\documentclass[aps,twocolumn,superscriptaddress]{revtex4-1}

\usepackage{bm}
\usepackage{gensymb} %for the degree symbol
\usepackage{graphicx} %eps figures can be used instead
\usepackage{amsmath} %for thhe matrix definition
\usepackage{breqn} %to break equations into multiple lines automatically, using the dmath environment
\usepackage{empheq,etoolbox} %for subequations
\usepackage{xargs}                      % Use more than one optional parameter in a new commands
\usepackage[pdftex,dvipsnames]{xcolor}  % Coloured text etc.
\usepackage{lipsum}
\usepackage{graphicx}
\usepackage{hyperref}
\usepackage[colorinlistoftodos,prependcaption]{todonotes}
\newcommandx{\unsure}[2][1=]{\todo[linecolor=red,backgroundcolor=red!25,bordercolor=red,#1]{#2}}

\newcommand{\Pe}{{\rm Pe}}

\newcommand{\Psimoy}{{\Psi}}
\newcommand{\Zdot}{{\dot{Z}}}

\makeatletter
\let\cat@comma@active\@empty
\makeatother

\begin{document}

\title{Rheotaxis of chiral bacteria: from single-cell behavior to a population-level description}
\author{Gustave Ronteix}
\affiliation{Physical microfluidics and Bioengineering, Institut Pasteur, 25-28 Rue du Dr Roux, 75015 Paris, France}
\affiliation{LadHyX, CNRS, Ecole Polytechnique, Institut Polytechnique de Paris, 91120 Palaiseau, France}
\author{Christophe Josserand}
\affiliation{LadHyX, CNRS, Ecole Polytechnique, Institut Polytechnique de Paris, 91120 Palaiseau, France}
\author{Ad\'ela\"{\i}de Lety-Stefanka}
\altaffiliation{Currently at Heriot-Watt University, Edinburgh, Scotland}
\affiliation{LadHyX, CNRS, Ecole Polytechnique, Institut Polytechnique de Paris, 91120 Palaiseau, France}
\author{Charles N. Baroud}
\affiliation{Physical microfluidics and Bioengineering, Institut Pasteur, 25-28 Rue du Dr Roux, 75015 Paris, France}
\affiliation{LadHyX, CNRS, Ecole Polytechnique, Institut Polytechnique de Paris, 91120 Palaiseau, France}
\author{Gabriel Amselem}
\email{amselem@ladhyx.polytechnique.fr}
\affiliation{LadHyX, CNRS, Ecole Polytechnique, Institut Polytechnique de Paris, 91120 Palaiseau, France}
\date{\today}

\begin{abstract}
    Due to their morphology, the dynamics of bacteria suspended in media can exhibit complex behaviors. In the presence of a shear, swimming bacteria experience a drift perpendicular to the shear plane. This drift, termed rheotaxis, is studied here semi-analytically, numerically and experimentally. We find the dependency of bacterial orientation and bacterial speed on the shear rate, in the presence of rotational diffusion. This enables us to show that the drift speed of bacteria perpendicular to the shear is predominantly due to bacterial propulsion, and not rheotactic forces. Comparing the drift speed of bacteria and diffusion leads to the definition of a P\'{e}clet number.  The rheotactic effect increases with the shear, reaching a plateau at very large P\'{e}clet numbers, in  good agreement with experiments of rheotaxis performed in microfluidic droplets.
    %The fraction of bacteria experiencing rheotaxis is found theoretically to increase with the P\'{e}clet number, reaching a plateau at very large P\'{e}clet numbers, in very good agreement with experiments of rheotaxis performed in microfluidic droplets. 
\end{abstract}

\maketitle

The motility of microorganisms is a crucial biological function, enabling these microorganisms to explore their environment and to migrate to more favorable habitats~\cite{Purcell1977LifeNumber, Lauga2009TheMicroorganisms, Kantsler2014RheotaxisCells}. Yet, many microorganisms such as bacteria or microalgae live in aqueous media, where their motility can be drastically affected by the surrounding fluid motion. The most obvious consequence of a surrounding flow on motility is that microorganisms are entrained by the flow. However, microorganisms can also travel perpendicular to the flow direction, due to the combined effects of fluid shear stress and of an  asymmetry in the system. 

The asymmetry can come from a directional stimulus. For example, the model microalgae \textit{Chlamydomonas reinhardtii} is known to experience gyrotaxis: when the shear due to a downwards Poiseuille flow couples to the action of gravity, algae are focused to the center of the flow~\cite{kessler1985}. Likewise, when \textit{C. reinhardtii} swimming in a Poiseuille flow are stimulated with a localized source of light, the combination of shear and phototaxis leads to their focusing to the centerline of the Poiseuille profile~\cite{garcia2013}. But the asymmetry can also come from the microorganism itself: when the helix-shaped spirochete \textit{Leptospira biflexa} is placed in a Poiseuille flow, it drifts perpendicular to the streamlines, a result of the interaction between shear and the chirality of the helix~\cite{marcos2009}. Motile bacteria in a Poiseuille flow likewise experience a lift force perpendicular to the plane of the shear~\cite{marcos2012}, a behavior termed rheotaxis and due to the interaction between three components: bacterial propulsion, shear, and the chiral geometry of the flagellar helix.

The seminal experiment on bacterial rheotaxis in a shear flow  was reported by Marcos et al. and consisted in flowing \textit{Bacillus subtilis} bacteria  in a Hele-Shaw microfluidic channel~\cite{marcos2012}. The shear profile of the Poiseuille flow in the channel led to bacterial rheotactic motion towards the side walls of the microchannel. The average rheotactic velocity was quantified experimentally as a function of shear rate, and experiments were in excellent agreement with numerical results. Mathijssen et al.~\cite{mathijssen2019} followed up on the study of bacterial rheotaxis and analyzed the motion of the bacteria \textit{Escherichia coli} in the vicinity of the solid walls of a microfluidic Hele-Shaw cell. They were able to observe and model four different regimes of bacterial swimming as a function of the applied shear rate. More recently, Jing et al.~\cite{Jing2020Chirality-inducedFlows} determined experimentally the orientation and rheotactic velocity of chiral bacteria in shear flows, comparing them to numerical results. In these three studies~\cite{marcos2012,mathijssen2019, Jing2020Chirality-inducedFlows}, the influence of the shear rate on the rheotactic velocity was obtained by numerically solving the equations for swimming at low-Reynolds number, and fitting them to functional forms.

Here, we  obtain semi-analytical formula for the rheotactic velocity of bacteria as a function of the shear rate. We begin by analyzing the case where bacteria follow purely deterministic trajectories in a shear flow. Bacterial trajectories are then the result of three compounding effects: (i) Jeffery orbits, due to the elongated shape of bacteria, which make the bacteria rotate in the plane of the shear~\cite{Jeffery1922TheFlow}; (ii) rheotaxis, due to the chirality of the bacterial flagella, that tends to both tilt bacteria out of the plane of shear, and translate them perpendicular to the plane of shear; and (iii) bacterial propulsion, which tends to make bacteria swim in a straight line. The combination of bacterial swimming and rheotaxis eventually makes the bacteria propel perpendicularly to the shear plane. In particular, we show that Jeffery effects and rheotaxis occur on two different time scales: the instantaneous dynamics of bacterial trajectories are dominated by Jeffery orbits, while the long-term trajectories show a net drift due to rheotaxis. We recover similar dynamics to that obtained analytically very recently by Ishimoto~\cite{ishimoto2020}.
 
In a second part, we take into account the effect of the randomization of bacterial trajectories, which can be characterized by an effective diffusion coefficient. While shear reorients bacteria perpendicularly to the shear plane, bacterial diffusion tends to randomize bacterial orientations.  The competition between diffusion and rheotaxis is captured in a P\'{e}clet number. At low P\'{e}clet numbers, diffusion dominates and all bacterial orientations are equiprobable. At high P\'{e}clet numbers, rheotactic reorientation dominates and bacteria tend to orient perpendicular to the plane of shear. The probability distribution of the angle of orientation of the bacteria is obtained numerically from the corresponding Fokker-Planck equation. Knowing this probability distribution enables us to calculate the average rheotactic reorientation in a population of bacteria as a function of the P\'{e}clet number, and so obtain the population-average speed of bacteria in the presence of rheotaxis and diffusion. 

In a confined experimental setup, bacteria experiencing a shear flow accumulate at the boundaries of the setup~\cite{Marcos2012BacterialRheotaxis,Jing2020Chirality-inducedFlows}. The proportion of bacteria accumulating at the boundaries is obtained from a Fokker-Planck equation, where randomization comes from rotational diffusion or from the run-and-tumble mechanism of bacteria, and where the average drift depends on the P\'{e}clet number. Theoretical results are in good agreement with experimental results of bacterial rheotaxis in microfluidic droplets, where the steady-state distribution of bacteria experiencing rheotaxis is easier to monitor than in standard open channels.

\section{Deterministic equations of motion}

\subsection{Definition of the problem}

\begin{figure*}[htb!]
\centering
 \center\includegraphics[width=15cm]{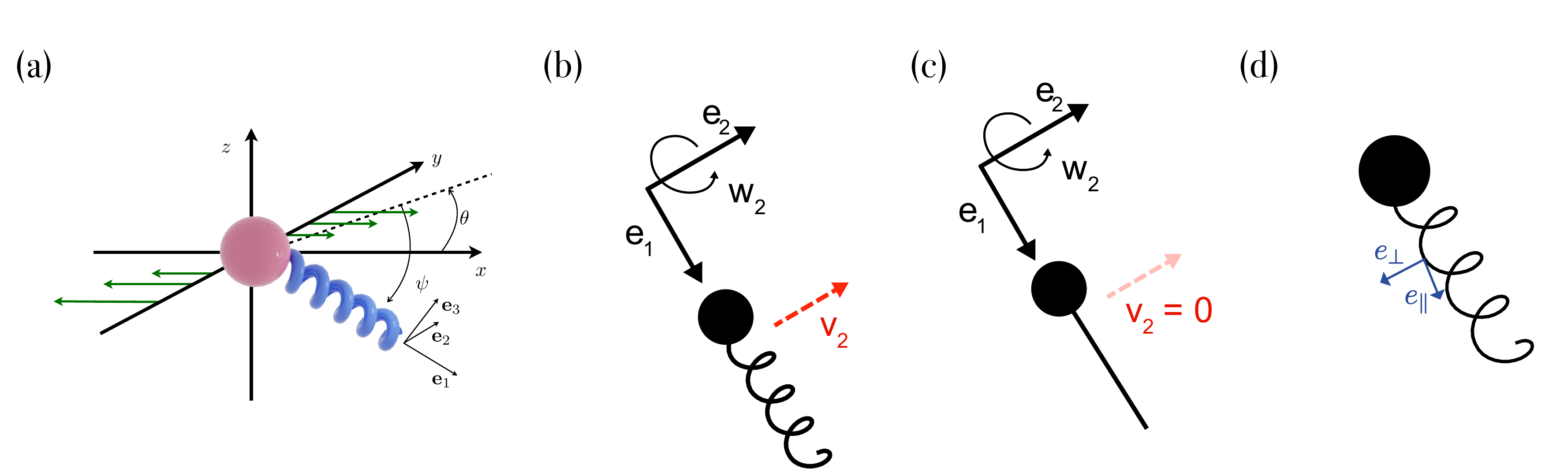} 
 \caption{(a) The bacteria is modeled as spherical head attached to a helicoidal flagellum. A shear is imposed in the $xy$-plane, so that the external fluid flow is $\mathbf{v_f}=\dot{\gamma}y\mathbf{e_x}$. The bacterial orientation is defined by the angle $\theta$ in the shear plane, and by the pitch angle $\psi$ perpendicular to the shear plane. The coordinate system attached to the bacteria is $(\mathbf{e_1}, \mathbf{e_2}, \mathbf{e_3})$, where  $\mathbf{e_1}$ is along the main axis of the bacteria. (b, c)  The  motion of the bacteria orthogonally to its main axis $\mathbf{e_1}$ is  coupled to its rotation through the term $B$ in the motility matrix $\zeta$. (b) A cell with a chiral helix will experience motion upon rotation. (c) For an achiral bacterial flagella, $B = 0$ and the rotation perpendicular to $\mathbf{e_1}$ doesn't induce any motion of the bacteria. (d) Definition of the orthogonal and parallel unit vectors $e_{\perp}$ and $e_{\parallel}$ along the flagellar helix. }
 \label{fig:figure_1_paper}
\end{figure*}

Rheotaxis originates from an interplay between bacterial chirality, bacterial motility, and the presence of a shear flow. A bacteria is modelled as a spherical head of radius $a$ attached to a helicoidal flagellum, see Fig.~\ref{fig:figure_1_paper}a. Call $R$ the radius of the flagellum,  $l$ its  total contour length, $\alpha$ its pitch, and $n$ the number of turns it makes. These geometrical parameters are related through $2\pi R n/l = \sin\alpha$. The location of a point on the helix is defined by its curvilinear coordinate $s$, with $0\leq s \leq l$. Last, a right-handed helix (resp. left) has handedness $\chi=1$ (resp. $\chi=-1$). 

The orientation of the bacteria in the lab frame is defined by the two angles $(\theta, \psi)$, where $\theta$ is the azimuthal angle of the bacteria in the $xy$-plane, and $\psi$ is the angle the bacteria makes with the $xy$ plane, see Fig.~\ref{fig:figure_1_paper}a. The center of the bacterial head is assumed to lie at the origin of the coordinate system.  An external shear flow is imposed in the $xy$-plane, such that the fluid velocity at any point $\mathbf{r} = (r, \theta, \psi)$ is given by  $\mathbf{v}_f(\mathbf{r})=\dot{\gamma}y\mathbf{e}_x = \dot{\gamma}r\cos\psi\sin\theta\mathbf{e}_x$, where $\dot{\gamma}$ is the shear rate and is defined by $\dot{\gamma} = \frac{\partial}{\partial y} \left( \mathbf{v}_f(\mathbf{r}) \cdot \mathbf{e}_x \right)$.

%%%%%%%%%

The Reynolds number associated to a bacteria of size $\approx 10\; \mu \rm m$ swimming at a speed  $\approx 30 \; \mu \rm m/s$ in water,  is ${\rm Re} \approx 3\times 10^{-4} \ll 1$. At this low Reynolds number, the force of the water on the spherical bacterial head is given by Stokes' formula:  $\mathbf{F}_{\rm head} = 6\pi\eta (\mathbf{v}_f(\mathbf{0}) - \mathbf{v}) = -6\pi\eta \mathbf{v}$, where $\eta$ is the viscosity of the surrounding fluid and  $\mathbf{v}$ is the speed of the center of mass of the bacteria, taken to be at the center of the spherical head~\cite{childress_1981}.

Resistive force theory allows us to calculate the viscous force on the slender bacterial helix, by decomposing the friction force locally on the helix into two components: one locally parallel to the helix orientation, and one perpendicular to it~\cite{childress_1981}. The two friction coefficients are $\zeta_{\parallel}$ and $\zeta_{\perp}$ respectively and are given by:
\begin{equation}
\zeta_{\parallel} = \frac{4\pi\eta}{\log\left(4\lambda^2/e^2\right) - 1} \quad {\rm and } \quad \zeta_{\perp} = \frac{8\pi\eta}{\log\left(4\lambda^2/e^2\right) + 1},
\end{equation}
where $\lambda$ is the wavelength of the helix and $e$ is the half-thickness of the helix~\cite{childress_1981}. At a point $P$ on the helix, defined by its coordinate $\mathbf{r}_P$, the helix velocity is $\mathbf{v}_{h}(\mathbf{r}_P)$ and the relative velocity between fluid and helix is $\mathbf{v}_{\rm rel}(\mathbf{r}_P) = \mathbf{v}_f(\mathbf{r}_P) - \mathbf{v}_{h}(\mathbf{r}_P)$. We define  $\mathbf{e}_{\parallel}(\mathbf{r}_P)$ and $\mathbf{e}_{\perp}(\mathbf{r}_P)$ the unit vectors that are tangent and perpendicular to the helix at point $P$. Resistive force theory gives the friction force locally: 
\begin{equation}
 \begin{split}
\mathbf{f}_{\rm helix}(\mathbf{r}_P)  = &  \zeta_\perp \left [\mathbf{v}_{\rm rel}(\mathbf{r}_P)\cdot \mathbf{e}_{\perp} \right ] \mathbf{e}_{\perp}+ \zeta_\parallel \left [\mathbf{v}_{\rm rel}(\mathbf{r}_P)\cdot \mathbf{e}_{\parallel} \right ] \mathbf{e}_{\parallel} \\
 = &  \zeta_\perp \mathbf{v}_{\rm rel}(\mathbf{r}_P) + (\zeta_\parallel - \zeta_\perp) \left [\mathbf{v}_{\rm rel}(\mathbf{r}_P)\cdot \mathbf{e}_{\parallel} \right ] \mathbf{e}_{\parallel}.
\end{split}   
\end{equation}
The total force on the helix is then $\mathbf{F}_{\rm helix} = \int_{0}^{l}\mathbf{f}(\mathbf{r}(s)) ds$. 

Last, let us calculate the total torque on the bacteria. Call $\bm{\omega}$ the rotation rate of the bacterial head, and $\mathbf{\Omega}_{\rm f}$ the curl of the fluid flow at the head center. The torque on the head is given by Faxen's second law:  $\mathbf{T}_{\rm head} = 8\pi\eta a^3 (\frac{\mathbf{\Omega}_{\rm f}}{2} - \bm{\omega}   )$~\cite{HildingFaxenDerIst}. The torque on the helix is  $\mathbf{T}_{\rm helix} = \int_{0}^{l}\mathbf{r}(s)\times \mathbf{f}(\mathbf{r}(s)) ds$. 

At low Reynolds number, inertia can be neglected, so that the sum of forces and sum of torques on the bacteria are zero at all times:

\begin{subequations}
  \begin{empheq}[left=\empheqlbrace]{align}
    & \mathbf{F}_{\rm head} + \mathbf{F}_{\rm helix}  = 0 \label{eq:zeroForce}  \\
    & \mathbf{T}_{\rm head} + \mathbf{T}_{\rm helix}  = 0 \label{eq:zeroTorque}
  \end{empheq}
  \label{eq:newton}
\end{subequations}

Equations~\ref{eq:zeroForce} and~\ref{eq:zeroTorque} are a linear system of 6 equations with 6 unknowns, which can be rewritten in matrix form as $\bm{\zeta}\mathbf{W} = \mathbf{A}$, where $\bm{\zeta}$ is the $6\times 6$ mobility matrix of the problem, $\mathbf{W} = (v_x,v_y, v_z, {\omega}_{x}, {\omega}_{y}, {\omega}_{z})$ is a column vector with all the unknowns that we want to solve for, and $\mathbf{A}$ is a forcing term with contributions from both the external flow and from the autonomous rotation of the helix~\cite{Marcos2012BacterialRheotaxis}.

This system proves extremely cumbersome to solve in the lab reference frame. Indeed, the helix orientation is characterized by 3 angles: the two angles defining the bacterial orientation $(\theta, \psi)$, as well as a third angle $\phi$ corresponding to rotation of the helix around its axis, see Fig.~\ref{fig:figure_1_paper}.  To go from the helix reference frame to the lab reference frame, one has to multiply the vectors $\mathbf{r}(s)$ and $\mathbf{e}_{\parallel}$ by the Euler rotation matrix $\mathbf{E} = R_{\phi}R_{\psi}R_{\theta}$, where 
$R_{\theta}  = \begin{pmatrix}
   \cos\theta & \sin\theta & 0\\
    -\sin\theta & \cos\theta & 0 \\
    0 & 0 & 1
  \end{pmatrix}$, 
$R_{\psi}  = \begin{pmatrix}
   \cos\psi & 0 & -\sin\psi\\
    0 & 1 & 0 \\
    \sin\psi & 0 & \cos\psi  

  \end{pmatrix}
$,
$R_{\phi}  = \begin{pmatrix}
    1 & 0 & 0 \\
    0 &\cos\phi & \sin\phi\\
    0 & -\sin\phi & \cos\phi 
    
  \end{pmatrix}$, and trigonometric terms then pervade the mobility matrix $\bm{\zeta}$.  

The trick is rather to solve the system of equations in the reference frame of the bacteria. We define the orthonormal basis $(\mathbf{e}_1, \mathbf{e}_2, \mathbf{e}_3)$, where $\mathbf{e}_1$ is along the bacterial axis, $\mathbf{e}_2$ is along  $\theta$, and $\mathbf{e}_3$ along $\psi$, see Fig.~\ref{fig:figure_1_paper}a.   The mobility matrix in the reference frame of the bacteria can then be calculated analytically and has a simple expression:

\begin{equation}
\label{eq:motilityMatrix}
\bm{\zeta} = \begin{pmatrix}
 A_f & 0 & 0 & M & 0 & 0\\
 0 & A & 0 & 0 & B & C \\
 0 & 0 & A & 0 & C & B\\
 M & 0 & 0 & F & 0 & 0\\
 0 & B & C & 0 & D & 0\\
 0 & C & B & 0 & 0 & D\\

\end{pmatrix},
\end{equation}
where the terms are given by:

\begin{dmath}
    A_f = -6\pi\eta a + l\zeta_{\parallel}\left(-1 + \frac{\zeta_{\perp}}{\zeta_{\parallel}}\sin\left(\alpha\right)^2\right)
\end{dmath}

\begin{dmath}
\label{coeff:M}
    M = \frac{1}{2\pi}l^2\zeta_{\parallel}\left(-1+\frac{\zeta_{\perp}}{\zeta_{\parallel}}\right)\cos\left(\alpha\right)\sin\left(\alpha\right)^2
\end{dmath}

\begin{dmath}
\label{coeff:A}
    A = -6\pi\eta a - \frac{1}{4}l\zeta_{\parallel}\left[1+3\frac{\zeta_{\perp}}{\zeta_{\parallel}} + \left( -1 + \frac{\zeta_{\perp}}{\zeta_{\parallel}} \right) \cos\left(2\alpha\right)\right]
\end{dmath}

\begin{dmath}
\label{coeff:B}
    B = \frac{1}{4\pi}l^2\zeta_{\parallel}\left(-1+\frac{\zeta_{\perp}}{\zeta_{\parallel}}\right) \cos\left(\alpha\right)\sin\left(\alpha\right)^2
\end{dmath}

\begin{dmath}
    C = \frac{1}{4}l^2\zeta_{\parallel}\cos\left(\alpha\right) \left[-2\frac{\zeta_{\perp}}{\zeta_{\parallel}} + \left( -1 + \frac{\zeta_{\perp}}{\zeta_{\parallel}} \right) \sin\left(\alpha\right) \right]^2
\end{dmath}

\begin{dmath}
    F = -8\pi\eta a^3 - \frac{1}{2}lR^2\zeta_{\parallel}\left[1+\frac{\zeta_{\perp}}{\zeta_{\parallel}} + \left( -1 + \frac{\zeta_{\perp}}{\zeta_{\parallel}} \right) \cos\left(2\alpha\right)\right]
\end{dmath}

\begin{dmath}
    D = -8\pi\eta a^3 - \frac{1}{24}l\zeta_{\parallel}\left[-12R^2\frac{\zeta_{\perp}}{\zeta_{\parallel}} + 4\left(3R^2\left( -1 + \frac{\zeta_{\perp}}{\zeta_{\parallel}} \right)-2l^2\frac{\zeta_{\perp}}{\zeta_{\parallel}}\right) \cos\left(\alpha\right)^2 + l^2\left( -1 + \frac{\zeta_{\perp}}{\zeta_{\parallel}} \right) \sin \left(2\alpha\right)^2\right]
\end{dmath}

%%%%%%%%%%%%
\subsection{Brief Observations}

The mobility matrix highlights key aspects of bacterial motility. The coupling between the bacteria rotation and the velocity along the bacteria's main axis $\mathbf{e}_1$ is given by the coefficient $M$. We can immediately verify that the bacteria does not undergo any motion when the helix is not chiral ($\alpha = 0$ or $R=0$), or in the absence of a helix ($l = 0$). The asymmetry in the friction coefficients is likewise crucial for motion at low Reynolds number: $M=0$ when $\zeta_{\parallel}=\zeta_{\perp}$.

For a chiral object, there are also couplings between translation and rotation along the two other principal axes: a rotation of the bacteria around $\mathbf{e}_2$ (resp. $\mathbf{e}_3$) leads to a translation around $\mathbf{e}_2$ (resp. $\mathbf{e}_3$), see Fig.~\ref{fig:figure_1_paper}b. This coupling is determined by the coefficient $B$ of the mobility matrix  (Eq.~\eqref{coeff:B}). %the coupling  between $\mathbf{e}_3$ and $\mathbf{\Omega}_3$, as well as between $\mathbf{e}_2$ and $\mathbf{\Omega}_2$, is determined by the coefficient $B$ in the motility matrix (Eq.~\eqref{coeff:B}). This coupling causes the motion of the bacteria out of the plane it is rotating in, leading to a net motion orthogonally to its rotation plane (Fig.~\ref{fig:figure_1_paper} b).
Conversely, for a radially symmetric object, $\alpha = 0$ and so $B = 0$: rotation and translation along $\mathbf{e}_2$ and $\mathbf{e}_3$ are not coupled (Fig.~\ref{fig:figure_1_paper}c).

We use Mathematica to solve the $6 \times 6$ system of equations~\eqref{eq:newton} in the reference frame of the bacteria $(\mathbf{e}_1, \mathbf{e}_2, \mathbf{e}_3)$, giving us access to the velocity and angular velocity of the bacterial head in this reference frame. Then, by returning to the laboratory frame we access the velocity and angular velocity of the bacterial head in the lab frame. In particular, we obtain the rheotactic velocity of the bacteria in the $z$ direction as a function of $\theta$ and $\psi$, coming from the shear on the helix in the $xy$ plane. 

The obtained expressions for the rheotactic velocity $\dot{z}$ as well as for the angular velocities $\dot{\theta}$ and $\dot{\psi}$  are extremely complex; they can be accessed in the Mathematica notebook in the Supplementary Material. Instead of manipulating them in their general form, we plug in numbers for the geometry of the bacteria, which allows us to obtain semi-analytical results for the rheotactic velocity as a function of the shear rate.

\section{Instantaneous, deterministic bacterial dynamics}
\label{sec:instantaneous}

\subsection{Plugging the bacterial geometric properties into the equations of motion}

Experimentally, the ratio between the parallel and perpendicular friction coefficients for a bacterial flagellum is $\zeta_{\parallel}/\zeta_{\perp}\approx 1.7$~\cite{marcos2012}. Considering a bacterial head with $a\approx 1 \; \mu {\rm m}$, a left-handed helix ($\chi = -1$) of length $l\approx 15 \; \mu \rm m$, half-thickness $e\approx 240 \; \rm nm$ and wavelength $\lambda\approx 3 \; \mu \rm m$, we find $\zeta_{\parallel} \approx 2\eta$, where    $\eta = 10^{-3} \; \rm Pa.s$ is the dynamic viscosity of water. The mobility matrix then contains several coefficients which are the sum of two terms of the same order of magnitude: $6\pi\eta a \approx 2\times 10^{-8}\; \rm kg.s^{-1}$, and  $l\zeta_{\parallel}\approx 3\times 10^{-8}\; \rm kg.s^{-1}$.

Replacing all geometric characteristics with their numerical values, we obtain long and cumbersome semi-analytical expressions for the evolution of the angular velocities and the rheotactic velocity. These expressions are rounded off to their two largest terms, giving us approximated equations for the angular velocities and the rheotactic velocity of the bacteria in the lab frame:

\begin{empheq}[left = \empheqlbrace]{align}
\dot{\theta}  =&  A_1 \dot{\gamma} + A_2 \dot{\gamma} \sin \left( \theta \right)^2 
    \label{eqn:thetaEq} \\
    \dot{\psi}   =& B_{\text{Jeff}} \dot{\gamma}  \sin \left( 2\psi \right) \sin \left( 2\theta \right)  \notag\\
      & \qquad + B_{\text{\text{rheo}}}\dot{\gamma} \cos \left( \psi \right) \cos \left( 2\theta \right)  \label{eqn:psiEq} \\
 \dot{z}   =& v_{\text{prop},z} +  C_{\text{Jeff}} \dot{\gamma} \sin \left( 2 \theta \right) \cos \left( \psi \right) \sin \left( 2\psi \right) \notag\\
    & \qquad + C_{\text{\text{rheo}}} \dot{\gamma} \cos \left( 2 \theta \right) \cos \left( \psi \right)^2.
    \label{eqn:zEq}
\end{empheq}

The constants $A_2$, $B_{\rm Jeff}$ and $C_{\rm Jeff}$ are non-zero only when the object studied is elongated. The constants $B_{\text{rheo}}$ and $C_{\text{rheo}}$ are non-zero when the object is chiral. Experimental values for the constants in Eq.~\eqref{eqn:thetaEq} -\eqref{eqn:zEq} are $A_1 \approx -0.02$, $A_2 \approx -0.94$, $B_{\text{Jeff}} \approx -0.94$, $B_{\text{\text{rheo}}} \approx -4 \cdot 10^{-3}$, $C_{\text{Jeff}} \approx -9 \cdot 10^{-7}$m and $C_{\text{\text{rheo}}} \approx 1.6 \cdot 10^{-8}$m. 

Studying these equations allows us to shed some light on the intricate dynamics of the bacteria. In the common case of an object with a radial symmetry axis (for instance an ellipsoid) subject to shear flow, the shear forces have the shear plane $(xOy)$ as a symmetry plane. These forces being supported by polar vectors, we expect the effects to be symmetric with regard to the  $(xOy)$ plane according to the Curie principle~\cite{Curie1894SurMagnetique}, and one can indeed verify that a radially symmetric object does not undergo any net motion in the $z$-direction. The addition of a chiral helix to the object adds a new rheotactic term in the equation, which breaks the symmetry of the problem, the helix itself not being symmetric with regard to the $(xOy)$ plane. In the following paragraphs we will study the effect of these rheotactic terms on the bacterial dynamics.

\begin{figure*}[htb!] 
\center\includegraphics[width=15cm]{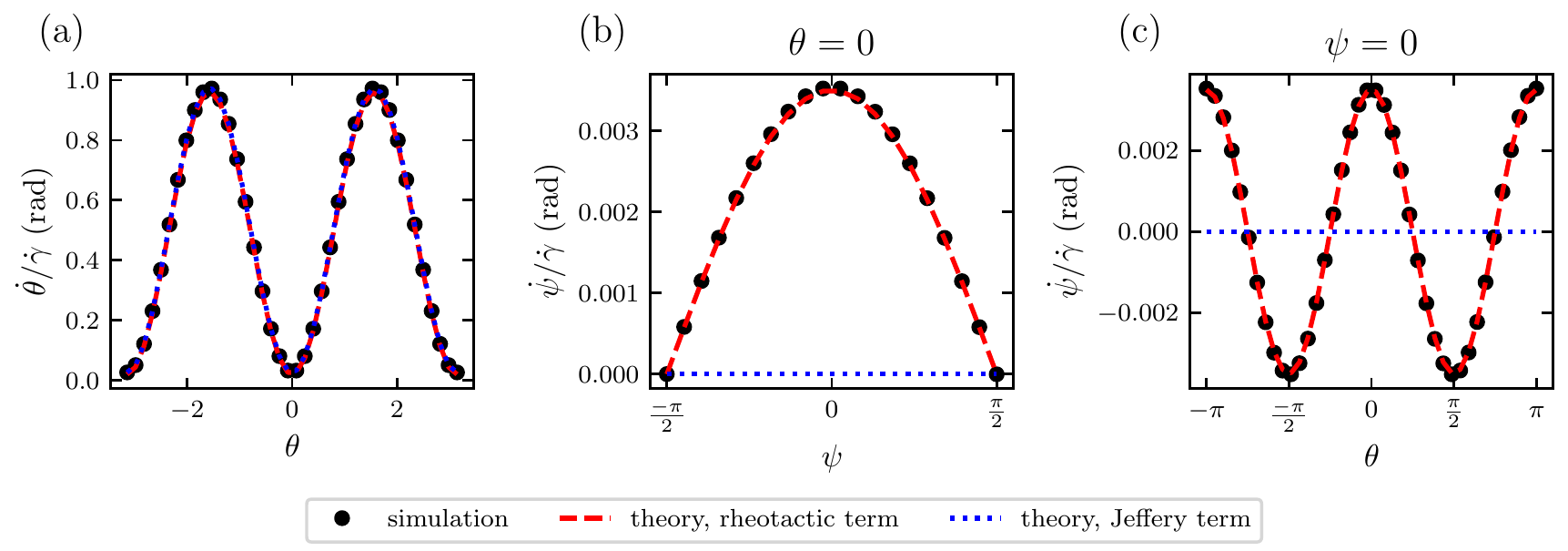}
\caption{Angular dynamics of bacteria with a right-handed helix in a shear flow. Black dots: results of numerical integration of the equations for bacterial motion as used in~\cite{Marcos2012BacterialRheotaxis,Mathijssen2019OscillatoryBacteria,Jing2020Chirality-inducedFlows}. Dashed red line: rheotactic contribution to $\dot{\psi}$, semi-analytical results . Dotted blue line: analytical result for an achiral elongated particle (Jeffery contribution). (a) The evolution of the angle in the shear plane $\theta$ is independent of the chirality of a particle and is given solely by Jeffery dynamics, see Eqn.~\eqref{eqn:thetaEq}.  (b) An achiral particle experiences no tilt force out of the shear plane ($\dot{\psi} = 0$, blue dotted line), in contrast to a bacteria (black dots and red dashed line). (c) An achiral particle in the shear plane stays in the shear plane (dotted blue line), while a chiral particle experiences a rheotactic tilt (black dots and red dashed line).}
\label{fig:numerical_validation_angles} 
\end{figure*}

The complexity of the motion of bacteria in shear flow  previously required the use of fully numerical methods to derive bacterial dynamics~\cite{Mathijssen2019OscillatoryBacteria, Jing2020Chirality-inducedFlows,Marcos2012BacterialRheotaxis}, or of a complex theoretical apparatus~\cite{ishimoto2020}. By changing the reference frame we have derived the corresponding semi-analytical equations. To check the approximated results in Eq.~\eqref{eqn:thetaEq}-\eqref{eqn:zEq}, we ran numerical simulations of the full equations of motion of bacteria in a shear flow. Simulation results enabled to retrieve the angular dynamics as well as the rheotactic force on bacteria.  Simulation results are in excellent agreement with the approximated equations~\eqref{eqn:thetaEq}-\eqref{eqn:zEq}, as detailed below (Figs. \ref{fig:numerical_validation_angles} and \ref{fig:numerical_validation_vz}). %in figures .

\subsection{Bacterial rotation in the shear plane $\theta$}

The dynamics of $\theta$ described by Eqn.~\eqref{eqn:thetaEq} are characteristic of the motion of an elongated object in a shear flow at low Reynolds number, first studied by Jeffery~\cite{jeffery1922}. In such a configuration, the object rotates in the plane of the shear, with a rotation rate proportional to the shear rate $\dot{\gamma}$. The constants $A_1$ and $A_2$ are related to the the effective aspect ratio $a_r=\sqrt{\frac{A_2+A_1}{A_1}}\approx 6.3$ of the bacteria.  Note that this rotation in the shear plane does not depend on the chirality of the object, but solely on its aspect ratio. Eq.~\eqref{eqn:thetaEq} can be rewritten as a function of $a_r$ and $\dot{\gamma}$~\cite{Leal1971TheFlow}:
\begin{dmath}
    \dot{\theta}  =  \frac{\dot{\gamma}}{1+a_r^2}\left( \cos \left( \theta \right)^2 + a_r^2 \sin \left( \theta \right)^2 \right),
    \label{eqn:thetaEqJeffery}
\end{dmath}
The evolution of $\theta$ is then periodic with a period $T = \frac{2\pi}{\dot{\gamma}}\left(a_r + \frac{1}{a_r}\right)$, and given by~\cite{Leal1971TheFlow}:
\begin{equation}
\tan \left(\theta\right) = a_r \tan \left(\frac{2 \pi t}{T} \right).
\label{eqn:thetaEqPeriod}
\end{equation}

Results of numerical simulations of bacterial motion  in a shear flow show excellent agreement with the analytical functional form given in Eqn.~\eqref{eqn:thetaEq}, see Fig.~\ref{fig:numerical_validation_angles}a. An elongated object spends most of its time aligned with the shear ($\theta = 0$ and $\theta = \pm \pi$), independently of its chirality, see Fig.~\ref{fig:numerical_validation_angles}a and~\cite{jeffery1922}.

\subsection{Instantaneous dynamics of the pitch angle $\psi$}

The dynamics of $\psi$, the angle made by the bacteria with the shear plane, result from two independent contributions: $\dot{\psi} = \dot{\psi}_{\text{Jeff}} + \dot{\psi}_{\text{\text{rheo}}}$, with $\dot{\psi}_{\text{Jeff}} =  B_{\text{Jeff}} \dot{\gamma}  \sin \left( 2\psi \right) \sin \left( 2\theta \right)$, and  $\dot{\psi}_{\text{\text{rheo}}} = B_{\text{\text{rheo}}}\dot{\gamma} \cos \left( \psi \right) \cos \left( 2\theta \right)$. The first term was already discussed by Jeffery~\cite{jeffery1922}, and stems from the interaction between shear and an elongated object. It depends solely on the aspect ratio of the elongated particle through the constant $B_{\rm Jeff}$. The second term in Eqn.~\eqref{eqn:psiEq} is due to the chirality of the helix, and causes the bacteria to tilt away from the shear plane. For a realistic bacterial geometry, the instantaneous dynamics of $\psi$ are dominated by Jeffery effects: $\dot{\psi}_{\text{Jeff}} \approx 0.94 \gg  \dot{\psi}_{\text{\text{rheo}}}\approx  4 \cdot 10^{-3}$.

The dependency of $\dot{\psi}$ on $\psi$ for a bacteria aligned with the shear ($\theta = 0$, the orientation in the shear plane in which bacteria spends most of its time) shows that the rheotactic term induces a bacterial tilt out of the shear plane, until the bacteria reaches an angle $\psi = \pi/2$, see Fig.~\ref{fig:numerical_validation_angles}b. This evolution is due to rheotactic effects: an achiral object making an angle $\psi$ with the shear plane, and aligned with the shear ($\theta=0$) does not have a tendency to be tilted out of the shear plane:  $\dot{\psi}_{\text{Jeff}}(\theta=0) = 0$, see blue dotted line in Fig.~\ref{fig:numerical_validation_angles}b. 

The angular dynamics of $\dot{\psi}$ also depend on the angle $\theta$ of the bacteria in the shear plane. They are plotted in the particular case of a bacteria in the shear plane ($\psi = 0$) in Fig.~\ref{fig:numerical_validation_angles}c: a bacteria in the shear plane gets tilted out of the shear plane because of rheotactic effects. This is in contrast with the evoution of an achiral object in the shear plane, for which $\dot{\psi}(\psi=0) = \dot{\psi}_{\text{Jeff}}(\psi=0) = 0$: an achiral object in the shear plane stays in the shear plane, see blue dotted line in Fig.~\ref{fig:numerical_validation_angles}c. 

\begin{figure*}[htb!] 
\center\includegraphics[width=0.9\textwidth]{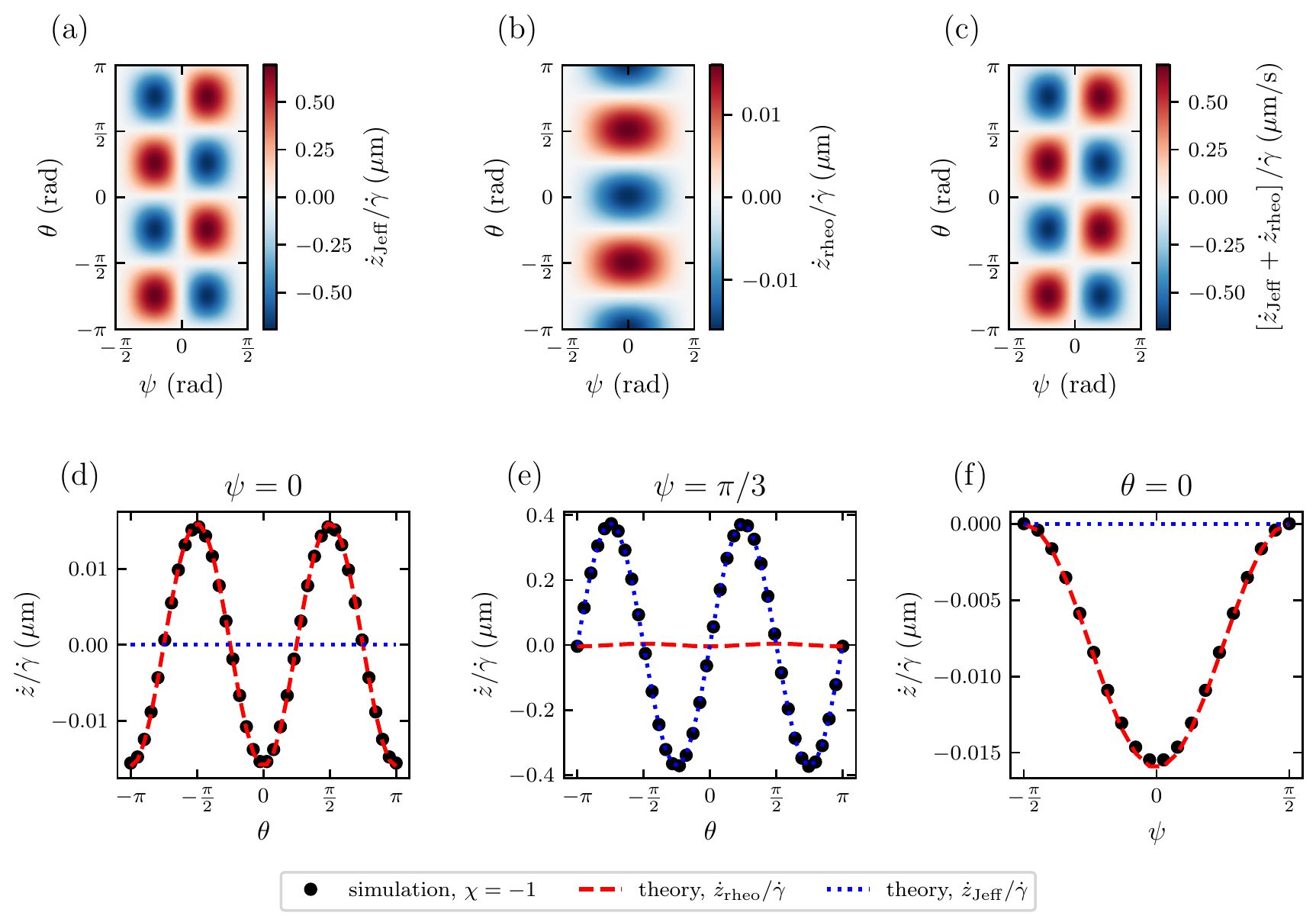}
\caption{Instantaneous velocity in $z$ for a bacteria with a right-handed helix (\textit{i.e.} $\chi = -1$), as a function of the angle in the shear plane $\theta$ and the tilt angle $\psi$. Black dots: numerical resolution of the bacterial dynamics. Dashed red line: semi-analytical result for the rheotactic contribution. Dotted blue line: Jeffery contribution. Velocity in the $z$ direction due to (a) Jeffery effects, (b) rheotactic effects as a function of $\left( \theta, \psi \right)$ . (c) Total instantaneous velocities in the $z$ direction. (d) At $\psi=0$, the drift is dominated by rheotactic effects (red dashed line), while Jeffery effects do not contribute (blue dotted line). (e) At $\psi=\pi/3$, the rheotactic drift is dominated by Jeffery effects (blue dotted line), while rheotactic effects are negligible (red dashed line). (f) A bacteria aligned with the shear ($\theta=0$) experiences a force that pushes it  along $-z$ depending symmetrically on the pitch angle $\psi$.}
\label{fig:numerical_validation_vz} 
\end{figure*}

\subsection{Instantaneous dynamics of the position $z$}

The dynamics of $z$ are described by Eq.~\eqref{eqn:zEq} and depend on three contributions: the rotation of the helical flagella, which propels the bacteria along $z$, $\dot{z}_{bac} = v_{\text{prop},z}$;  a term due to Jeffery orbits $\dot{z}_{\text{Jeff}} = C_{\text{Jeff}} \dot{\gamma} \sin \left( 2 \theta \right) \cos \left( \psi \right) \sin \left( 2\psi \right)$ (Fig.~\ref{fig:numerical_validation_vz}a); and a term due to the chirality of the helix $\dot{z}_{\text{\text{rheo}}} = C_{\text{\text{rheo}}} \dot{\gamma} \cos \left( 2 \theta \right) \cos \left( \psi \right)^2$ (Fig.~\ref{fig:numerical_validation_vz}b). To clarify the interplay between the Jeffery and rheotactic effects, we assume for the time being that the bacteria are not self-propelled: $v_{\text{prop},z} = 0$. Note that $C_{\text{Jeff}} \gg  C_{\text{\text{rheo}}}$, and so for most angles $(\theta, \psi)$, the instantaneous dynamics of $z$ are dominated by Jeffery effects (Fig.~\ref{fig:numerical_validation_vz}c). 

The relative contributions of the Jeffery and rheotactic effects on the instantaneous speed $\dot{z}$ are plotted in the $(\theta, \psi)$ plane in Fig.~\ref{fig:numerical_validation_vz}a-c. When $\theta \approx 0~[\pi/2]$, and when $\psi\approx 0$, the velocity in $z$ is predominantly due to  rheotactic effects. % In all other cases however, the $z$-velocity is caused by Jeffery effects, see e.g. Fig.~\ref{fig:numerical_validation_vz}e for $\psi=\pi/3$. 
For all other values of $(\theta, \psi)$ however, the bacteria drifts in $z$ mainly due to Jeffery effects. Indeed, the maximum drift speed along $z$ due to rheotaxis is one order of magnitude smaller than that due to Jeffery effects, see Fig.~\ref{fig:numerical_validation_vz}a,b. 

Three particular cases of the evolution of the drift speed $\dot{z}$ with the angles $\theta$ and $\psi$ are highlighted in Fig.~\ref{fig:numerical_validation_vz}d-f, for a bacteria (points and red dashed line) and for an achiral elongated particle (blue dotted line). A bacteria in the shear plane ($\psi = 0$) displays a drift in $z$ solely due to rheotaxis, and whose direction depends on the orientation $\theta$ of the particle with respect to the shear direction; an achiral elongated particle in the shear plane does not drift in $z$, see Fig.~\ref{fig:numerical_validation_vz}d. A bacteria tilted with respect to the shear plane ($\psi \neq 0$) drifts mainly due to Jeffery effects, see Fig.~\ref{fig:numerical_validation_vz}e for $\psi = \pi/3$. A bacteria aligned with the shear direction ($\theta = 0$) drifts along $z$ due to rheotactic effects, while an achiral elongated particle does not drift, see Fig.~\ref{fig:numerical_validation_vz}f.

To summarize, the instantaneous drift in $z$ is dominated by rheotactic effects for $\psi\approx 0$ and by Jeffery effects otherwise. The maximum absolute rheotactic drift occurs for $\psi = 0$ and $\theta = 0~[\pi/2]$ see Fig.~\ref{fig:numerical_validation_vz}b.

\section{Long-term, deterministic bacterial dynamics}
\label{sec:longterm}

The instantaneous bacterial dynamics tend to show that dynamics are dominated by Jeffery effects. Yet, Jeffery effects do not lead to an overall net motion in the plane perpendicular to the shear~\cite{Hinch1979RotationFlow}. This contrasts with experimental results of bacterial rheotaxis in a shear flow~\cite{Marcos2012BacterialRheotaxis, Mathijssen2019OscillatoryBacteria, Jing2020Chirality-inducedFlows}, which  demonstrate a net movement of bacteria orthogonal to the shear plane. This apparent contradiction can be overcome by noting that, during a rotation of period $T$ of the bacteria in the shear plane, the Jeffery terms cancel out: $ \int_0^T \dot{\psi}_{\text{Jeff}} dt = 0$, but that the -- smaller in magnitude -- rheotactic terms do not: $ | \int_0^T \dot{\psi}_{\text{\text{rheo}}} dt | > 0$, see Eq.~\eqref{eqn:psiEq}. Indeed, the rheotactic forces exerted on the bacterial flagella induce a torque on the bacteria, leading to a progressive tilt in the pitch angle $\psi$ (Fig.~\ref{fig:figure_traj}a). The evolution of the bacteria in $z$ is thereby linked to its evolution in $\psi$, as demonstrated by numerical integration of the instantaneous equations of motion, which reveal a net drift in $\psi$ and $z$, see Fig.~\ref{fig:figure_traj}c-f. This net drift occurs on time scales longer than $T$.

\subsection{Long-term dynamics of $\theta$ and $\psi$}

The angle $\theta$ in the shear plane follows a periodic trajectory of period $T$ which does not depend on the chirality of the object in the shear flow, see Fig.~\ref{fig:figure_traj}b and Eq.~\eqref{eqn:thetaEq}: an elongated object essentially spends most of its time aligned with the shear direction, and very short periods of time rotating a full circle on itself. The pitch angle $\psi$ also shows oscillations with time, with a period $T/2$, see Fig.~\ref{fig:figure_traj}c and Eq.~\eqref{eqn:psiEq}. However, the trajectory of $\psi$ depends on the chirality $\chi$ of the object: for a chiral particle ($\chi \neq 0$), the value of $\psi$ displays a net drift superimposed to its oscillations (see blue line in Fig.~\ref{fig:figure_traj}c), while the trajectory of an achiral elongated particle ($\chi = 0$) does not display any drift (see red line in Fig.~\ref{fig:figure_traj}c and~\cite{Hinch1979RotationFlow}). 

The value of the pitch angle $\psi$ is averaged over several periods of rotation $T$ to get rid of the oscillations due to Jeffery effects and highlight its long-term drift. Such a moving average of the pitch angle $\psi$ over three periods of rotation $T$ is shown in Fig.~\ref{fig:figure_traj}e. The precise dynamics of the pitch angle $\psi$ depend on the initial value of the angle $\psi_0$, and in all cases the time-averaged pitch angle converges to a steady-state value $\pm\pi/2$. For a chiral object in a shear flow with a rate $\dot{\gamma}$ such that $\chi\dot{\gamma}>0$, the steady-state value is $ \pi/2$ (see Fig.~\ref{fig:figure_traj}e). When $\chi\dot{\gamma}<0$, the steady-state value is $ -\pi/2$. The instantaneous pitch angle shows oscillations around this steady-state value~\cite{Jing2020Chirality-inducedFlows}.

\subsection{Long-term dynamics of the position $z$}

\begin{figure*}[htb!] 
\center\includegraphics[width=\textwidth]{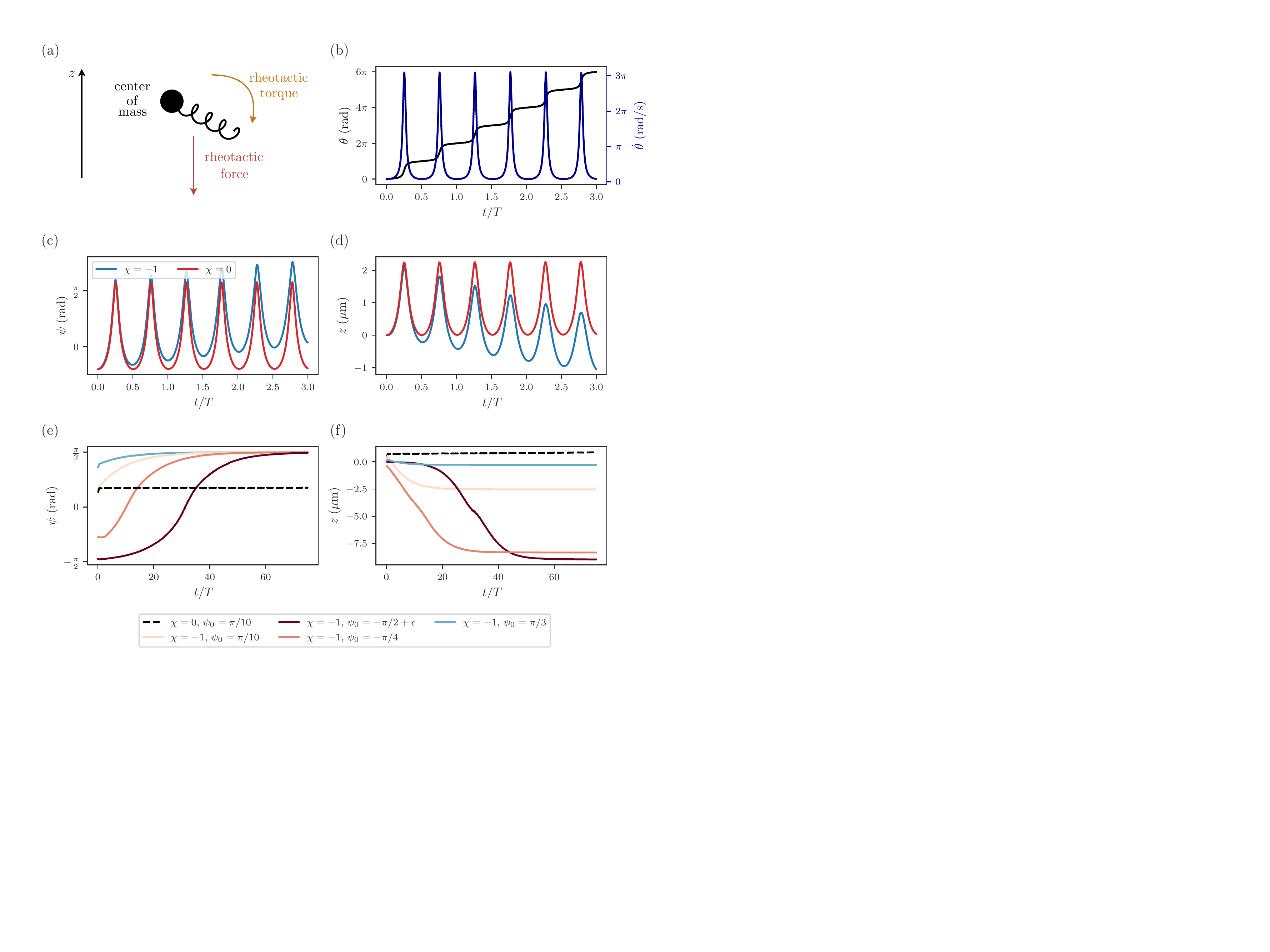}
\caption{Trajectories of a bacteria in a shear flow, in the absence of bacterial propulsion. (a) The rheotactic force is exerted on the bacterial flagella, causing the bacteria to move in the $z$ direction. The rheotactic force also exerts a torque, tilting the bacteria and aligning it with the $z$-axis. (b) The evolution of $\theta$ and $\dot{\theta}$ is periodic with a period $T$ given by Jeffery orbits. (c) Pitch angle $\psi$ of the bacteria as a function of time and of the chirality of the helix for an initial position $(\theta_0, \psi_0) = (0, \pi/10)$. Achiral bacteria (red line, $\chi = 0$) experience periodic variations of the pitch angle but no average drift, as expected for particles undergoing Jeffery orbits. In contrast, chiral bacteria (blue line, $\chi \neq 0$) experience quasi-periodic variations of the pitch angle with a long-term drift. (d) The  rheotactic force causes a long-term drift of the bacteria in $z$, in addition to the periodic Jeffery oscillations (blue line). An achiral particle shows solely periodic oscillations of $z$ (red line). (e) Long-term evolution of the pitch angle $\psi$, averaged over 3 periods of oscillation $T$, to highlight the rheotactic drift. $\psi$ converges to the stable point $\psi_{\inf} = \pi/2$ for left-handed chiral bacteria. (f) Average position of the bacteria $z$ (averaged over 3 periods of oscillation $T$).  For a left-handed bacteria in a shear flow with a positive shear rate $\dot{\gamma}$, the position $z$ decreases before stabilizing to a steady-state value that depends on the original orientation of the bacteria.}
\label{fig:figure_traj} 
\end{figure*}

The position $z$ of the particle likewise shows oscillations with a period $T/2$, for both chiral ($\chi \neq 0$) and achiral ($\chi = 0$) elongated particles, see Fig.~\ref{fig:figure_traj}d. The introduction of chirality leads to an overall drift of the particle along $z$, perpendicular to the shear plane, and on a time scale longer than $T$, see Fig.~\ref{fig:figure_traj}d. As for the pitch angle, this long-term drift in $z$ is better seen by averaging $z$ over three periods of rotation $T$, to get rid of the oscillations due to Jeffery orbits. The position $z$ evolves until stabilizing at a value that depends on the value of the initial angle $\psi_0$, see Fig.~\ref{fig:figure_traj}f. This stable position is reached when the chiral particle has stabilized itself at an angle $\psi = \pi/2$. There are then no more rheotactic effects:  Eq.~\eqref{eqn:psiEq} and~\eqref{eqn:zEq} show that $\dot{\psi} \sim \dot{\gamma} \cos \psi$ and $\dot{z} \sim \dot{\gamma} \cos^2 \psi$, both of which are 0 for $\psi = \pm \pi/2$. 

The drift along $z$ is coupled to the evolution of $\psi$, so that the distance traveled in $z$ depends on the time needed for $\psi$ to reach its stable-state $\psi=\pi/2$: the further away the initial value $\psi_0$ is from $\pi/2$, the longer the time to reach the stead-state in $\psi$ and therefore the further the drift in $z$, see Fig.~\ref{fig:figure_traj}f. This difference between chiral and achiral particles can also be visualized in the phase space $(\widetilde{x}, \widetilde{y}, \widetilde{z})$, where $\widetilde{x} = \cos (\psi) \cos (\theta)$, $\widetilde{y} = \cos (\psi) \sin (\theta)$, and $\widetilde{z} = \sin (\psi)$, showing periodic oscillations for an achiral particle, and a trajectory converging towards a fixed point for a chiral particle (see SI, Fig.~\ref{fig:supplementary_rheotaxis}).

Note that the dynamics described here reflect the behavior of non-motile bacteria, for which $v_{\text{prop},z} = 0$. A non-motile bacteria drifts along $z$ until it is tilted perpendicular to the shear plane, and has by then drifted to a stable $z$ position less than $10~\mu$m away from its starting position (Fig.~\ref{fig:figure_traj}f).  Adding self-propulsion leads to a long-term drift along the axis of the bacteria, which is perpendicular ($\psi=\pi/2$) to the shear plane, so that bacteria in a Poiseuille flow drift all the way to the channel sides~\cite{Marcos2012BacterialRheotaxis,Mathijssen2019OscillatoryBacteria,Jing2020Chirality-inducedFlows}.

\section{Randomization of bacterial trajectories by diffusion}

In the previous sections, bacterial propulsion was ignored, and all bacteria were found to eventually orient perpendicularly to the shear plane, as soon as they experience a non-zero shear rate. This however does not take into account two sources of noise in bacterial trajectories: bacteria are small enough to experience rotational diffusion, and motile bacterial display a run-and-tumble motion that leads to an effective randomization of their trajectories. In this section, we study the competition between the rheotactic force, which tends to align bacteria perpendicularly to the shear plane, and bacterial motility and diffusion, which tends to randomize bacterial orientation.

\subsection{The effect of rotational diffusion}

The rotational diffusion coefficient of bacteria is given by the Stokes-Einstein relationship $\mathcal{D}_{SE} = k_B T/D\approx 1.7 \cdot 10^{-3} \; \rm rad^2.s^{-1}$, where $k_B = 1.38 \times 10^{-23} \; \rm SI$ is Boltzmann's constant, $T= 293 \; \rm K$ is the ambiant temperature, and $D$ is the last diagonal term of the mobility matrix $\bm{\zeta}$, see Eq.~\eqref{eq:motilityMatrix}~\cite{Leal1971TheFlow}. This rotational diffusion coefficient depends solely on the geometry of the bacteria, and is independent of bacterial motility.

The  distribution $\rho({\theta,\psi,t})$ of the angles $(\theta,\psi)$ then satisfies the following Fokker-Planck equation~\cite{Leal1971TheFlow}:
\begin{equation}
\label{eqn:FP_psitilde}
    \frac{\partial \rho}{\partial t} = 
     \mathcal{D}_R\Delta\rho -\nabla\cdot(\bm{\dot{\omega}}\rho) , %\Delta \rho({\psi}) + \nabla \left(\dot{{\psi}(\psi, \theta)} \rho({\psi})\right),
\end{equation} 
where $\bm{\dot{\omega}} = (\dot{\theta}, \dot{\psi})$ is the angular velocity of the bacteria. 

From equations~\eqref{eqn:thetaEq} and~\eqref{eqn:psiEq}, we see that the Jeffery terms are significantly larger than the terms associated with the rheotactic force. This suggests that over short time-spans (typically of the order of a single rotational period $T$), the rotational dynamics is dominated by Jeffery effects. However, the long-term dynamics of the bacteria are caused by the interaction between the rheotactic force and diffusion. To investigate the long-term dynamics we introduce the average pitch angle $\psi$ over a period $T$: $\Psimoy = \frac{1}{T}\int_0^T \psi(t) dt$. The evolution of $\Psimoy$ is driven by the following Fokker-Planck equation:
\begin{equation}
\label{eqn:psitildeFP}
    0 = \mathcal{D}_{R} \Delta p - \nabla \cdot \left(\dot{\Psimoy} p\right),
\end{equation} where $p$ is the steady state distribution of $\Psimoy$. 

\begin{figure*}[htb!] 
\center\includegraphics[width=\linewidth]{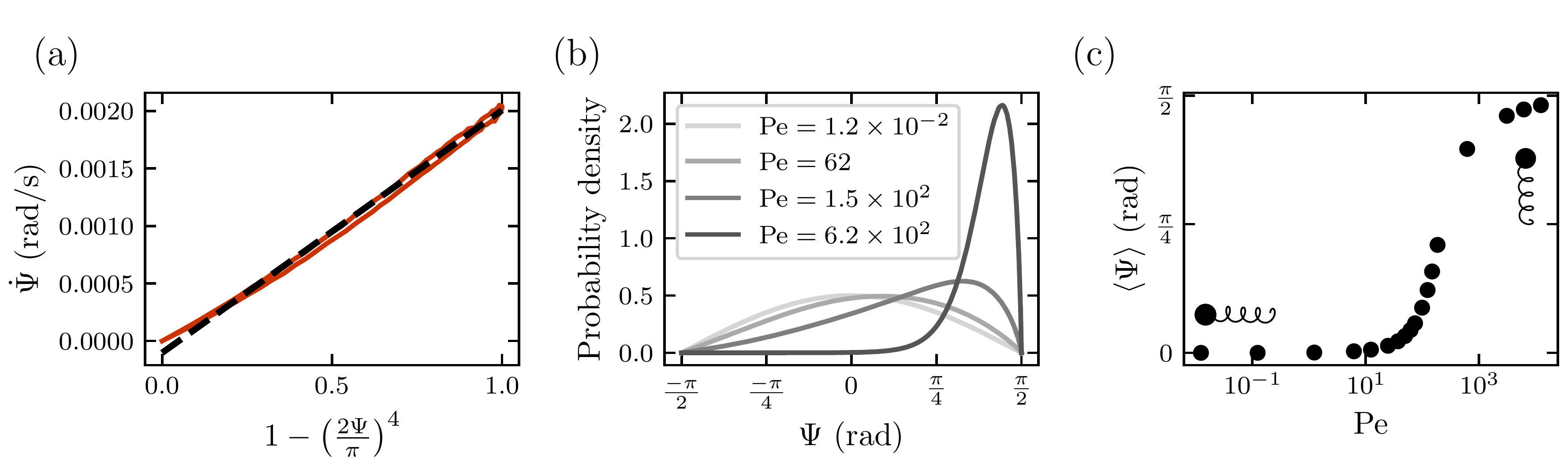}
\caption{(a) We can approximate the evolution of $\dot{\Psimoy}$ as a function of $\Psimoy$. The closer $\Psimoy$ is to $ \pm \pi/2$, the smaller the value of the rheotactic-driven forcing of the pitch angle. (b) The stronger the shear rate, the more the distribution of $\Psimoy$ is peaked close to $\Psimoy = \pi/2$. (c) Population average evolution of $\Psimoy$, $\langle \Psimoy \rangle$ as a function of shear stress. For low values of $\dot{\gamma}$, the bacteria are on average aligned with the shear plane. However, as the shear increases the bacteria orient themselves orthogonally to the shear plane.}
\label{fig:num_psi} 
\end{figure*}

It is necessary to know the dependency of $\dot{\Psimoy}$ on $\Psimoy$ to solve Eq.~\eqref{eqn:psitildeFP}. This dependency can be determined numerically through Eq.~\eqref{eqn:psiEq}, and is plotted in Fig.~\ref{fig:num_psi}a (orange line). It can be empirically fitted to the following function:
\begin{equation}
\label{eqn:psidotmean}
    \dot{\Psimoy} = a_{\Psimoy} \dot{\gamma} \left[ 1 - \left( \frac{2 \Psimoy}{\pi} \right)^4 \right],
\end{equation} where $a_{\Psimoy} = 2.1 \cdot 10^{-3} ~ \text{rad}$ (black dotted line in Fig.~\ref{fig:num_psi}a). The Fokker-Planck equation for $p(\dot{\Psimoy})$ is then numerically solved as a function of the  effective P\'{e}clet number $\text{Pe}_{\Psimoy} = \dot{\gamma} a_{\Psimoy}/\mathcal{D}_{R}$, giving the probability density to find bacteria at any given angle $\Psimoy$ (Fig.\ref{fig:num_psi}b). 

For small values of $\text{Pe}_{\Psimoy}$ (corresponding to shear values of typically $\dot{\gamma} = \mathcal{O} \left( 1 \right)~ \rm s^{-1} $), the equilibrium distribution is close to that expected for randomly distributed particles:  $p\left( \Psimoy \right) \sim \frac{1}{2}\cos \Psimoy$, see light gray lines in Fig.\ref{fig:num_psi}b. Increasing the shear rate leads to a sharper accumulation of the bacterial orientation toward $\pi/2$, see darker lines in Fig.\ref{fig:num_psi}b. This competition between rheotactic and randomizing effects can be visualized by defining the population average of $\Psimoy$: $\langle \Psimoy \rangle = \int \Psimoy p\left( \Psimoy \right) d\Psimoy$. For low values of the shear rate, bacteria are oriented in all directions and the population-averaged $\langle \Psimoy \rangle$ is 0. As the shear rate is increased, bacteria become on average more oriented perpendicularly to the shear plane, and $\langle \Psimoy \rangle$ converges toward $\pi/2$, see Fig.~\ref{fig:num_psi}c.

Rheotactic forces and diffusion together shape the equilibrium distribution of the bacterial tilt. This enables us to compute the velocity orthogonally to the shear plane by including both the rheotactic force and bacterial propulsion.

%%%%
\subsection{Retrieving the drift in $z$}
\label{sec:gammatovel}

The bacterial drift in the $z$ direction, perpendicular to the shear plane, depends on the intrinsic bacterial propulsion along $z$, as well as on the rheotactic drift. We define $\Zdot = \frac{1}{T} \int_0^T \dot{z} (t)dt$, the average velocity in $z$ over a single rotational period of the bacteria. Equation~\eqref{eqn:zEq} shows that $\Zdot$ is the sum of two contributions: the rheotactic contribution $\Zdot_{\text{rheo}}$, and the propulsion contribution $\Zdot_{\text{prop}}$. The Jeffery contribution is null: an achiral elongated particle does not drift along $z$.

\subsubsection{Drift due to rheotaxis}

Numerical integration of the deterministic equations of motion shows that the average rheotactic drift during a period $T$ can be empirically fitted to $\Zdot_{\text{rheo}} \approx a_{\dot{z}} \dot{\gamma} \cos \left(\Psimoy \right)^2$, see Fig. \ref{fig:v_gamma}a. This expression can be used to calculate the population-averaged rheotactic drift $\langle \Zdot_{\text{rheo}} \rangle= \int \Zdot_{\rm rheo} p(\Psimoy) d\Psimoy$, where $p(\Psimoy)$ is the probability distribution of $\Psimoy$.

The population-averaged rheotactic drift $\langle \Zdot_{\text{rheo}} \rangle$ depends on the shear rate $\dot{\gamma}$ both through the expression of $\Zdot_{\text{rheo}}$ and through the evolution of $p(\Psimoy)$ with the shear. This combined dependency leads to a non-monotonic evolution of $\langle \Zdot_{\text{rheo}} \rangle$ with the shear. On the one hand, the higher the shear rate, the stronger the deterministic rheotactic drift, as $\Zdot_{\text{rheo}} \propto \dot{\gamma}$. On the other hand, the higher the shear rate, the more the distribution of $\Psimoy$ peaks close to $\pi/2$, which strongly reduces the rheotactic force, since $\Zdot_{\text{rheo}} \propto \cos \left(\Psimoy \right)^2$. This trade-off yields an optimum shear maximizing the rheotactic drift, see Fig.~\ref{fig:v_gamma}b.

\subsubsection{Drift due to bacterial propulsion}

\begin{figure*}[htb!] 
\center\includegraphics[width=\linewidth]{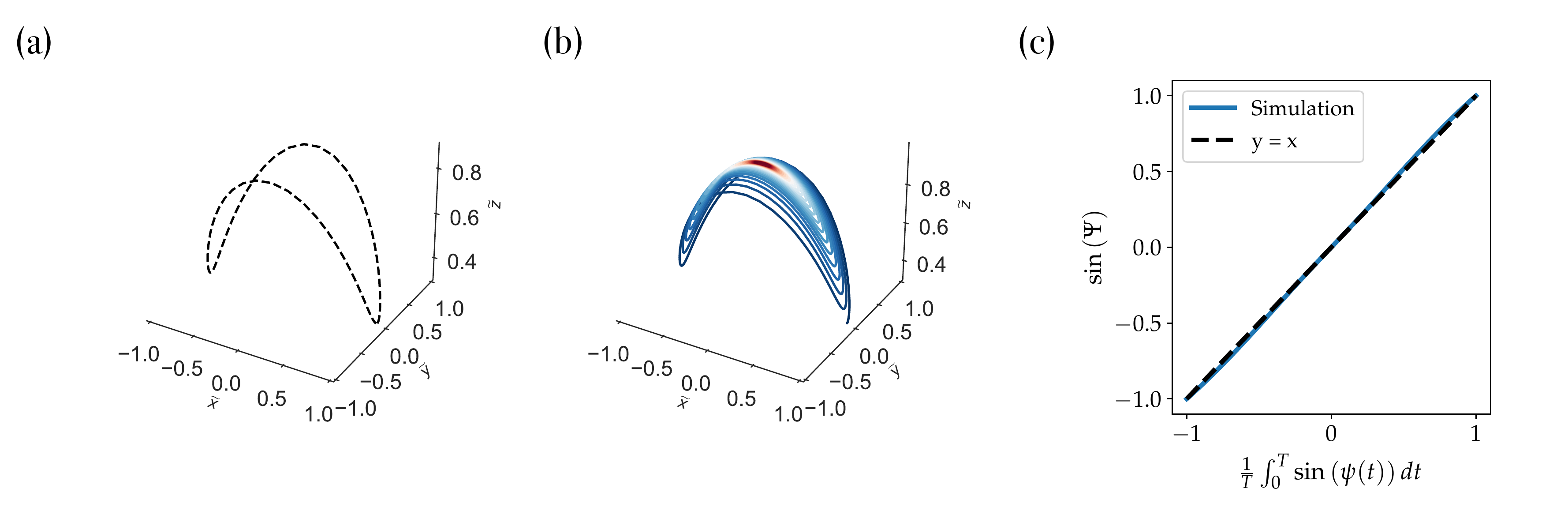}
\caption{(a) Trajectory of an achiral particle in shear flow on the unit sphere from a starting position $\left( \theta_0, \psi_0 \right) = \left( 0, \pi/10 \right)$. The particle oscillates periodically following Jeffery orbits. (b) In contrast, the rheotactic force exerted on a chiral bacteria causes a long term drift, biasing the trajectories. (c) The value of $\frac{1}{T} \int_0^T \sin \left( \psi (t) \right) dt$ fits closely to that of $\sin \left( \Psi \right)$, where $\psi(t)$ is given by Eqn.~\eqref{eqn:psiEq}.}
\label{fig:supplementary_rheotaxis} 
\end{figure*}

The average propulsion velocity perpendicularly to the shear plane is $\Zdot_{\text{prop}} = v_{\text{prop}} \frac{1}{T} \int_0^T \sin \left( \psi (t) \right) dt$, where $\psi(t)$ is given by equation~\eqref{eqn:psiEq}. Numerical integration shows that $\frac{1}{T} \int_0^T \sin \left( \psi (t) \right) dt \approx \sin \Psimoy$ (see SI, Fig.~\ref{fig:supplementary_rheotaxis}). We therefore obtain that, for a single bacterium, the propulsion velocity along $z$, averaged over a period of rotation $T$, is $\Zdot_{\text{prop}} \approx v_{\text{prop}} \sin \Psimoy$. This enables us to calculate the population averaged propulsion velocity $\langle \Zdot_{\text{prop}} \rangle$. At the population level, at low shear rates all bacteria are oriented randomly and $\langle \sin \Psimoy \rangle = 0$, so the population averaged propulsion velocity is $\langle \Zdot_{\text{prop}} \rangle = 0$. When the shear rate is increased, more and more bacteria orient perpendicularly to the shear plane, $\langle \sin \Psimoy \rangle$ converges to 1, and the population averaged propulsion velocity converges to $\langle \Zdot_{\text{prop}} \rangle = v_{\text{prop}}$, see Fig.~\ref{fig:v_gamma}c. 

\subsubsection{Comparison between drift due to rheotactic forces and to bacterial propulsion}

\begin{figure*}[htb!] 
\center\includegraphics[width=\linewidth]{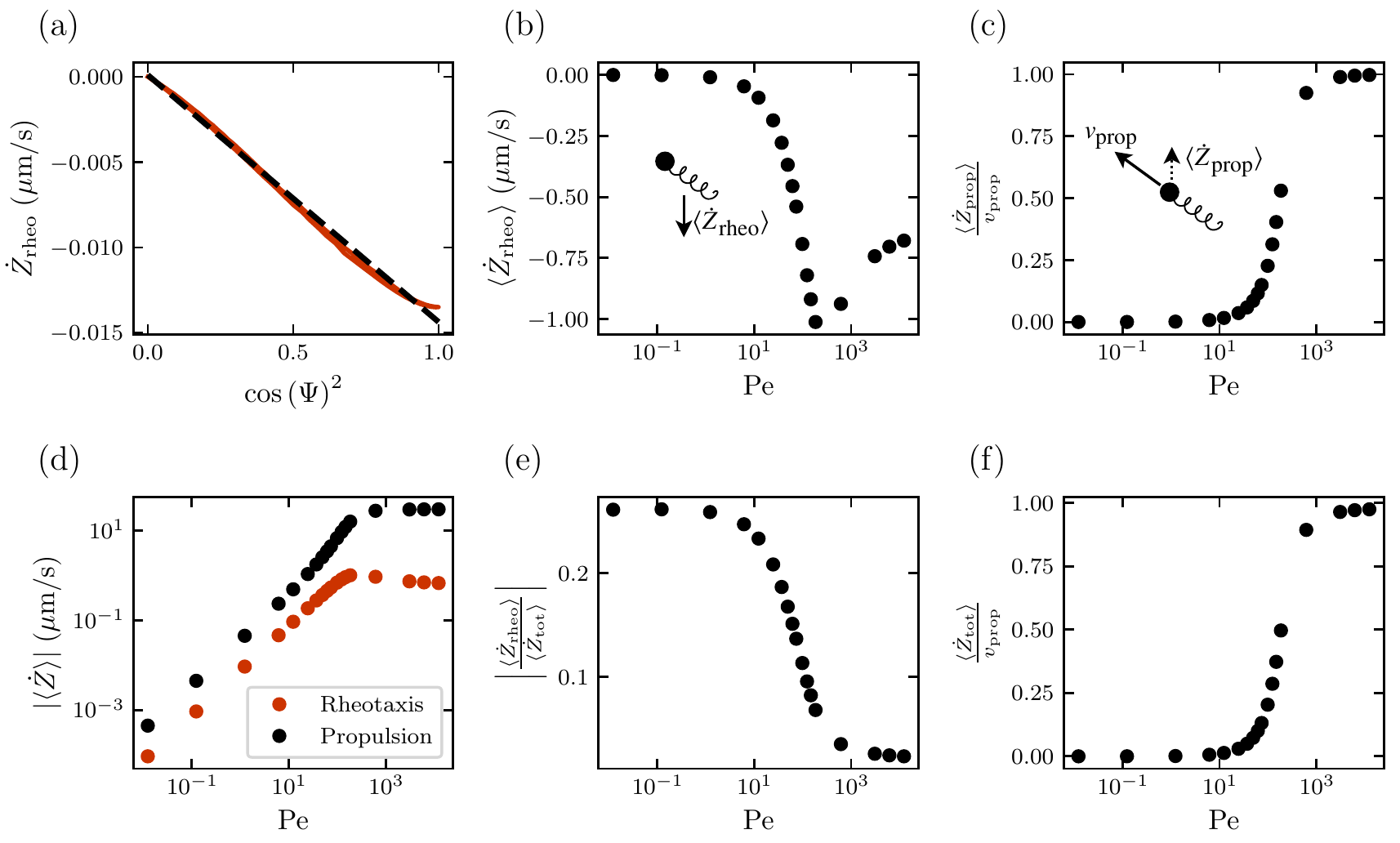}
\caption{(a) $\Zdot_{\rm rheo}$ depends linearly on $\cos \left( \Psimoy \right)^2$. (b) The value of $\langle \Zdot_{\rm rheo} \rangle$ depends non-monotonously on the value of the shear strain $\dot{\gamma}$: there exists a shear value $\dot{\gamma}_{\rm max}$ that maximizes the rheotactic drift of the bacteria. (c) Normalized bacterial velocity along the $z$ axis as a function of $\dot{\gamma}$. (d) Comparison of the motion orthogonal to the shear plane as a function of the shear due to rheotaxis (red dots) and propulsion (black dots). The motion in the $z$ direction is dominated by the self-propulsion of the bacteria. (e) The relative influence of the rheotactic drift on the total motion along the $z$ axis decreases for increasing values of $\dot{\gamma}$. (f) Conversely, the ratio between total displacement in $z$, $\langle \Zdot_{\rm tot} \rangle$ and the bacterial propulsion velocity increases with shear.}
\label{fig:v_gamma} 
\end{figure*}

What is the main cause underlying the drift of a bacterial population along $z$? For the realistic bacterial geometry considered here, this drift is always dominated by bacterial propulsion, and not by the anti-symmetric rheotactic force exerted on the chiral helix by the fluid, see Fig.~\ref{fig:v_gamma}d: the main effect of the rheotactic force is to reorient the bacteria perpendicularly to the shear plane, and bacterial propulsion then causes the bacteria to drift perpendicular to the shear plane. 

As the bacteria progressively reorient with stronger shear rates, the proportion of propulsion  attributable to the rheotactic drift decreases from about 25\% to less than 5\%, see Fig.~\ref{fig:v_gamma}e. This means that for high values of $\dot{\gamma}$, nearly all the velocity orthogonally to the shear plane is caused by the re-orientation and bacterial propulsion, see Fig.~\ref{fig:v_gamma}f.

\subsection{Diffusion in $z$}

The rotational diffusion of bacteria and their run-and-tumble mode of motion effectively leads to a randomization of their trajectories, which can be characterized by a translational diffusion coefficient. Here, we ask how bacteria are distributed along $z$, the axis perpendicular to the plane of shear. Calling $\mathcal{D}_z$ the effective diffusion coefficient in $z$, the  probability $p_z(z)$ for a bacteria to be at position $z$ obeys the following Fokker-Planck equation: 

\begin{equation}
\label{eqn:master}
    \frac{\partial p_z}{\partial t} = \mathcal{D}_{z} \Delta p_z - \nabla \cdot \left( \dot{z} p_z \right).
\end{equation}

As before, we consider the dynamics of bacteria averaged over one Jeffery orbit, and replace $\dot{z}$ with $\langle \Zdot \rangle = \langle \Zdot_{\text{prop}} \rangle + \langle \Zdot_{\text{rheo}} \rangle$ in equation~\eqref{eqn:master}. This allows us to find the steady-state probability distribution along the $z$-axis as a function of $\dot{\gamma}$, assuming a system of size $L$ along $z$, with Neumann boundary conditions.

\begin{equation}
\label{eqn:analyticalsolution}
    p_z(z) = \frac{\Pe}{L} \frac{e^{z\Pe/L}}{e^{\Pe} - 1},
\end{equation}
%where $L$ is the size of the experimental system along $z$, and 
where we have introduced the characteristic P\'{e}clet number $\Pe = L \langle \Zdot_{\text{tot}} \rangle /\mathcal{D}_{z}$ as the key parameter defining the bacteria probability density along the $z$-axis. When $\Pe \ll 1$, diffusion dominates rheotaxis and we can verify that the bacteria are uniformly distributed long $z$: $p_z(z) \sim \frac{1}{L}$. On the other hand, if $\Pe \gg 1$, then $p_z(z) \sim \delta (z,L)$ and we see that the probability distribution peaks at $z = L$: all bacteria experience rheotactic drift and accumulate at the boundary. The P\'{e}clet number depends on the shear rate $\dot{\gamma}$ through the average rheotactic speed $\langle \Zdot_{\text{tot}} \rangle$ which itself varies with shear as shown in section \ref{sec:gammatovel}.

\section{Experimental results}
\label{sec:experimental}

\begin{figure}[ht!]
\centering
 \includegraphics[width=0.5\textwidth]{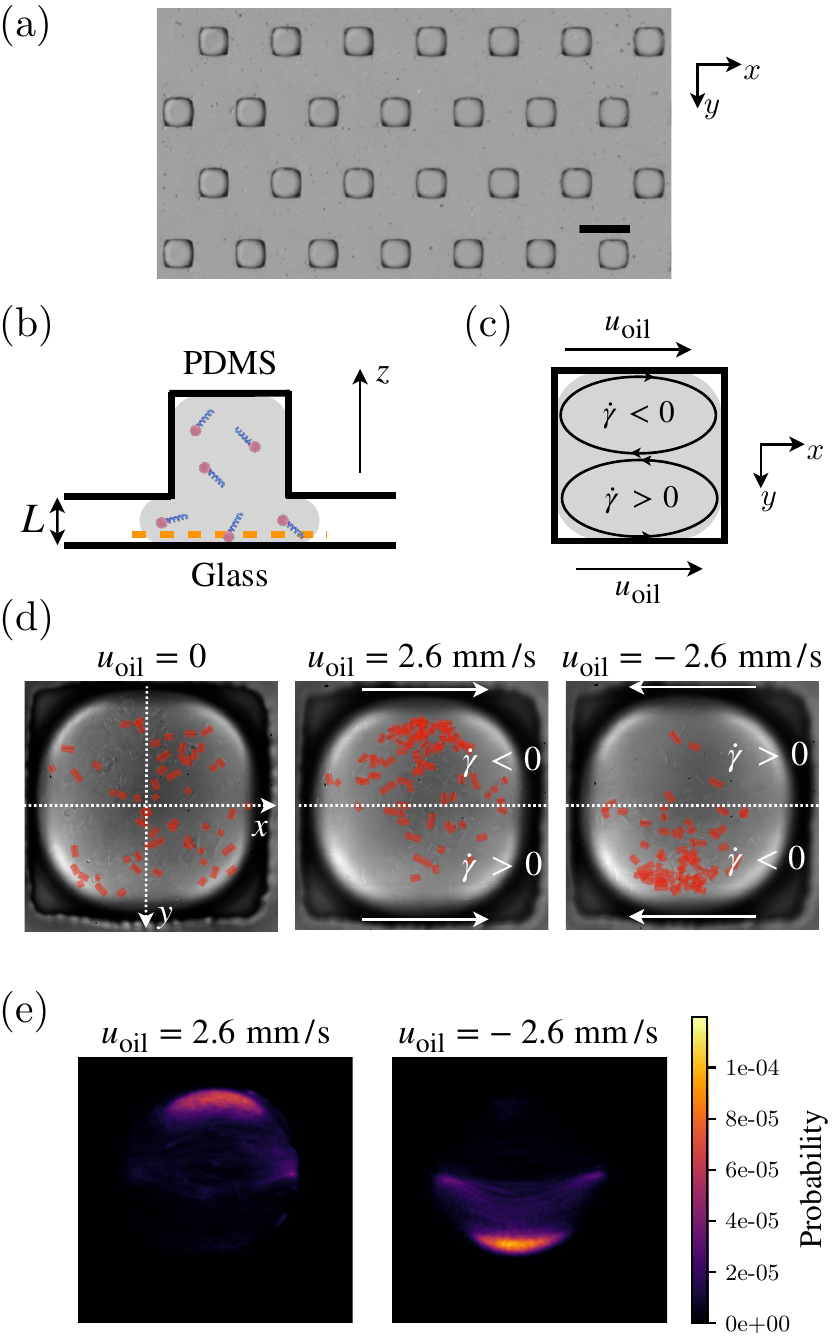} 
 \caption{(a) Snapshot of the microfluidic device (bottom-view). The squares are holes in the channel ceiling, that act as surface-energy anchors in which droplets are created and immobilized. Scale bar: $200\; \rm \mu m$. (b)  Cross-sectional schematic of an individual droplet immobilized in its anchors, containing bacteria and surrounded by oil. The plane of observation is shown by the orange dashed line. (c) Schematic of the flow in a trapped water droplet (gray): the outer oil flow leads to two recirculation regions in the  droplet, where the shear rate has opposite signs. (d) Left: Bacteria in a droplet in the absence of flow are randomly distributed . Middle: in the presence of an oil flow going from left to right ($u_{\rm oil} = 2.6 \; \rm mm/s$), recirculation within the aqueous drops induces bacterial segregation in the top-half ($y>0$) of the droplet. Right: in the presence of an oil flow going from right to left, bacteria segregate to the bottom-half ($y<0$) of the droplet  ($u_{\rm oil} = 2.6 \; \rm mm/s$). White arrows indicate the oil flow direction. Bacteria are highlighted in red for better visualization. (e) Probability density maps of bacterial presence, for an oil flow going from left to right ($u_{\rm oil} = 2.6 \; \rm mm/s$), and from right to left ($u_{\rm oil} = -2.6 \; \rm mm/s$).}
 \label{fig:device}
\end{figure}

The fraction of bacteria undergoing rheotaxis as a function of shear was experimentally investigated using a droplet microfluidics setup. Bacteria were encapsulated in microfluidic droplets of radius $R\approx 75 \;  \mu \rm m$, which were immobilized in square grooves etched in the microfluidic channel ceiling. The grooves,  of side $150 \;{\mu \rm m}$ and depth $100 \; \mu \rm m$, acted as surface-energy anchors~\cite{Dangla2011TrappingEnergy,Amselem2016UniversalDroplets}. Each droplet filled entirely its anchor, and bulged out in the main channel of height $L=35\; \mu \rm m$, see Fig.~\ref{fig:device}a-b.  Imposing an oil flow in the microfluidic chamber led to a symmetric recirculation pattern within the immobilized droplets, see  Fig.~\ref{fig:device}c and~\cite{Lee2012}. 
 
In the absence of an outer oil flow, bacteria swam freely within their droplets at a speed $v_{\text{prop}}\approx 30 \; \mu \rm m.s^{-1}$, and the distribution of bacteria in a droplet was uniform, see the first snapshot in Fig.~\ref{fig:device}d. When an outer oil flow was applied, the spatial distribution of bacteria within the droplets changed drastically and became asymmetric: within 5 seconds after imposing the  flow, bacteria segregated to one side of the droplet, see the middle snapshot in Fig.~\ref{fig:device}d for an average shear rate within the droplet $\dot{\gamma} \approx 2.8\; \rm s^{-1}$. Reversing the flow direction led to a reversal of the bacterial segregation pattern, see the last snapshot in Fig.~\ref{fig:device}d for an average shear rate within the droplet $\dot{\gamma} \approx -2.8\; \rm s^{-1}$. Probability density maps of the location of bacteria in the droplet highlight the segregation pattern in the presence of flow, with all bacteria being found in the top half (resp. bottom half) of the droplet for a shear rate $\dot{\gamma} \approx 2.8\; \rm s^{-1}$ (resp.  $\dot{\gamma} \approx - 2.8\; \rm s^{-1}$), see Fig.~\ref{fig:device}e.

The flow pattern within an anchored droplet or within a droplet moving in a Hele-Shaw cell is complex and three-dimensional~\cite{Lee2012,ling2016}. As a first approximation, the recirculation can be assumed to exist solely in the part of the droplet bulging out of the groove, in the main channel of height $L= 35\; \mu \rm m$. The recirculation can be neglected in the part of the droplet that is in the groove (data not shown). As a further simplification, for $0\leq z \leq L$, the flow in the droplet was assumed to be two dimensional, in the $xy$ plane, and such that the shear rate $\dot{\gamma}$ is equal and opposite in the two halves of the droplet, as indicated  in Fig.~\ref{fig:device}c~\cite{Lee2012,ling2016}.

The bacterial asymmetry was quantified by calculating the difference in number of bacteria between top and bottom half of the droplet, and normalizing it by the total number of bacteria: calling $N_{+}$ the number of bacteria in the top half of the droplet ($y>0$) and $N_{-}$ the number of bacteria in the bottom half of the droplet ($y<0$), the asymmetry was defined as $\alpha  = \frac{N_+ - N_-}{N_+ + N_-}$. The asymmetry is then $\alpha=-1$ when all bacteria are in the lower half of the droplet, $\alpha=+1$ when they are all in the top half of the droplet, and $\alpha=0$ when bacteria are at random positions in the droplet. 

\begin{figure}[ht!]
\centering
 \includegraphics[width=0.5\textwidth]{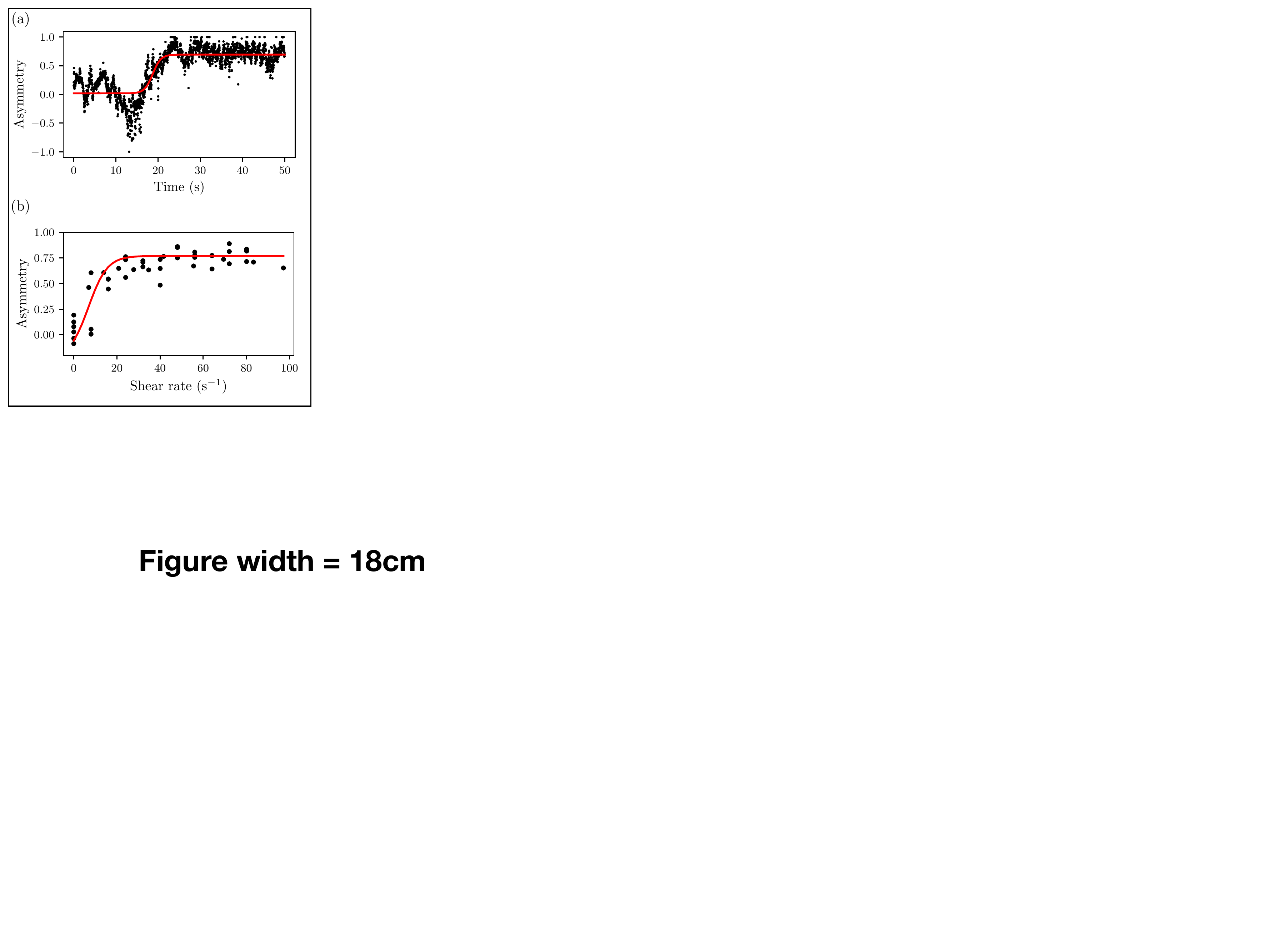} 
 \caption{ (a) Asymmetric segregation of  \textit{B. subtilis} in a droplet. At the beginning of the experiment, bacteria swim freely in the droplet and there is no asymmetry. At t=10~s, an outer oil flow of mean speed $u_{\rm oil} = -2.6 \; \rm mm/s$ is applied, leading to bacterial segregation in the droplet. Black points: experimental data. Red line: fit to a sigmoid function. (b) The end-value of the asymmetry depends on the applied shear. Black points: experimental data. Red line: fit to a sigmoid function enabling the extraction of an effective P\'eclet coefficient.}
 \label{fig:bacteria}
\end{figure}

In the absence of an oil flow ($t<12 \; \rm s$), the asymmetry fluctuated around 0.  Once the oil flow was applied, the asymmetry increased within $10\; \rm s$ to its steady-state value. An example of the evolution of the asymmetry with time is shown in Fig.~\ref{fig:bacteria}a, where the  mean oil speed is $u_{{\rm oil}} \approx 210\;  \mu {\rm m.s^{-1}}$, corresponding to a shear rate within the droplet of $\dot{\gamma} \approx u_{\rm oil}/R \approx 2.8\; \rm s^{-1}$. The steady-state value of the asymmetry was then $\alpha \approx 0.75$. The steady-state value of the asymmetry depended on the oil flow rate: the faster the oil speed $u_{\rm oil}$, the greater the shear rate $\dot{\gamma} \approx u_{\rm oil}/R$ in the droplet, and the greater the steady-state bacterial asymmetry, until reaching a plateau, see Fig.~\ref{fig:bacteria}a. The asymmetry plateaus at $\alpha = 0.75 < 1$ because of the presence of bacteria close to the centerline of the droplet that are entrained by secondary flows. 

To compare experimental and theoretical results, note that the two halves of the droplet have equal and opposite shear rates $\dot{\gamma}$. When the outer oil flowed from left to right, the P\'{e}clet number was therefore proportional to $\dot{\gamma}$ in the top half of the droplet ($y>0$), and to $-\dot{\gamma}$ in the bottom half of the droplet ($y<0$). As a first approximation, the flow speed along $x$ in the droplet is given by $v_x = \dot{\gamma} (|y| - R/2)$. Calling ${\Pe} = \langle \Zdot_{\text{tot}} \rangle L/\mathcal{D}_{z}$, the probability of finding a bacteria at a height $z$ in the top-half ($y>0$) of the droplet is then:

\begin{equation}
    p_{y>0}(z) = \frac{\rm Pe}{L}\frac{e^{z {\rm Pe}/L}}{e^{\rm Pe}-1}
\end{equation}

and the probability of finding a bacteria at a height $z$ in the bottom-half of the droplet is:
\begin{equation}
    p_{y<0}(z) = \frac{\rm Pe}{L}\frac{e^{-z{\rm Pe}/L}}{1-e^{-\rm Pe}}
\end{equation}

Calling $\Delta h$ the depth of field of the microscope, the average asymmetry is then given by:

\begin{align}
    \left\langle \frac{N_+-N_-}{N_+ + N_-} \right\rangle & = \frac{\int_{L-\Delta h}^{L}p_{y>0}(z)dz - \int_{L-\Delta h}^{L}p_{y<0}(z)dz}{\int_{L-\Delta h}^{L}p_{y>0}(z)dz + \int_{L-\Delta h}^{L}p_{y<0}(z)dz} \\
    & \approx \frac{p_{y>0}(L)-p_{y<0}(L)}{p_{y>0}(L) + p_{y<0}(L)}  \\
    & = \frac{e^{{\rm Pe}} + e^{-\rm Pe} - 2}{e^{\rm Pe} - e^{-\rm Pe}} 
    \label{eq:fracBacteriaVsPe}
    %& = \frac{\Delta h L\dot{z}}{D_0}\frac{e^{-hL\dot{z}/D_0}}{1-e^{-L\dot{z}/D_0}}.
\end{align}

A fit of the experimental data to Eqn.~\eqref{eq:fracBacteriaVsPe} is shown in Fig.~\ref{fig:bacteria}b. The P\'{e}clet number is given by ${\Pe} = \langle \Zdot_{\text{tot}} \rangle L/\mathcal{D}_{z}$, where $ \langle \Zdot_{\text{tot}} \rangle$ was previously calculated and $L=35\; \mu {\rm m}$ is the main channel height, in which the recirculation exists. The fit in figure~\ref{fig:bacteria} gives the relationship between the P\'{e}clet number and shear stress: ${\Pe} = \beta \dot{\gamma}$, where $\beta$ is derived from the fit. Combining these two expressions, we can derive an effective value of the diffusion coefficient $\mathcal{D}_{z} \approx 220\; \mu \rm m^2/s$, a value in the range of the effective diffusion coefficient of a bacteria executing a run-and-tumble motion ($\approx 100\mu \rm m^2/s$~\cite{berg1993random}). 
 
The estimation of $\langle \Zdot_{\text{tot}} \rangle$ is dependent on numerous geometric parameters where slight differences between experimental and theoretical values compound to yield significant differences. Furthermore, the flow is three-dimensional and more complex than assumed, so that the simple linear estimation of the shear rate used above is not correct. Nevertheless, despite these caveats, the estimated value of the diffusion coefficient is close to the values reported in the literature. 

%%%%%%%%%
%%%%%%%%%%%

%%%%%%%%%
%%%%%%%%%%%

\section{Summary and outlook}

Starting from the equations of motion of bacteria in a shear flow, we obtained semi-analytic formulas for the deterministic dynamics of the angles $\theta$ and $\psi$ describing the orientation of bacteria in and out of the shear plane, as well as for the evolution of $z$, the bacteria position perpendicular to the shear plane. We then considered the effect of noise on the bacterial dynamics, and retrieved the probability distribution at equilibrium for the pitch angle as well as for Theoretical results on the  fraction of bacteria displaying a rheotactic drift at  a given shear rate were favorably compared to experimental data of bacterial rheotaxis in microfluidic droplets.

Our theoretical results emphasize the two different time scales at play in bacterial rheotaxis: bacteria are elongated objects following Jeffery orbits in the shear plane, while at the same time experiencing a drift perpendicular to the shear plane (rheotaxis) due to their chiral helix. The instantaneous dynamics of bacterial trajectories are dominated by Jeffery effects (section~\ref{sec:instantaneous}), so that the angles that describe the bacterial orientation in the shear plane, $\theta$, and perpendicular to the shear plane, $\psi$, as well as the position $z$ of the bacteria all oscillate with a period given by Jeffery (see Eqn.~\ref{eqn:thetaEqJeffery} and Fig.~\ref{fig:figure_traj}b,c)~\cite{Jeffery1922TheFlow,Leal1971TheFlow}. Jeffery effects alone however do not lead to a drift perpendicular to the shear plane, which is instead due to the chirality of the bacterial helix and occurs on a longer time scale.

The evolution of the angle $\theta$ in the shear plane is not affected by rheotactic effects, see Eqn.~\eqref{eqn:thetaEq}. The dynamics of the pitch angle $\psi$ yet depends on the presence of the chiral helix, and the angle $\psi$ slowly drifts, over a time scale much larger than the period of oscillations $T$, until reaching the fixed point $\psi = \pi/2$ (see Fig.~\ref{fig:figure_traj}d). The rheotactic drift in $z$, perpendicular to the shear plane, depends on the pitch angle through $\dot{z}\propto\cos(\psi)$, and stops evolving for $\psi = \pi/2$ (see Fig.~\ref{fig:figure_traj}f). Once $\psi = \pi/2$ is reached, the  drift observed experimentally in the $z$ direction is purely due to intrinsic bacterial propulsion: bacteria are stably oriented perpendicularly to the shear, and move along their axis of orientation. 

The above picture is purely deterministic, and does not take into account the fact that bacteria are subject to rotational diffusion and exhibit a run-and-tumble motion that leads to a randomization of their trajectories~\cite{berg1993random,Saragosti2012ModelingChemotaxis}, thereby countering the rheotactic effects. To determine whether rheotaxis or diffusion dominates, the P\'{e}clet number ${\rm Pe} = a_{\psi}\dot{\gamma}/\mathcal{D}_R$ was introduced, where $\dot{\gamma}$ is the shear rate and $\mathcal{D}_R$ the effective rotational diffusion coefficient. The value of the numerical prefactor $a_{\psi} = 2.1 \cdot 10^{-3}\; \rm rad^2$ depends on the geometrical parameters of the bacteria and is determined from the deterministic equations of motion. At low values of the P\'{e}clet number, bacterial orientations are random, while at high values of the  the P\'{e}clet number, the distribution of the tilt angle is skewed towards $\pi/2$, see Fig.~\ref{fig:num_psi}b,c.  

The drift of bacteria along $z$, perpendicularly to the shear plane, is then the sum of two components: the rheotactic drift existing for all chiral objects, and intrinsic bacterial propulsion along the axis of the bacteria. The average value of each of these drift speeds at the population level depends on the average bacterial orientation, and so on the value of the P\'{e}clet number. When diffusion dominates, all bacterial orientations are equiprobable and as many bacteria experience a rheotactic drift along $z$ as along $-z$, so the average rheotactic speed is zero. When diffusion is negligible, all bacteria orient at $\pi/2$, an angle at which there is no rheotactic force, so that the rheotactic drift is also zero.  In between these two extremes, there exists a value of the P\'{e}clet number that leads to a maximum rheotactic speed, see Fig.~\ref{fig:v_gamma}b. For realistic bacterial geometries, the drift speed in $z$ due bacterial propulsion dominates that due to rheotaxis, for all values of the P\'{e}clet number, see Fig.~\ref{fig:v_gamma}d,e.

The rotational diffusion of motile bacteria, as well as their run-and-tumble motion lead to an effective translational diffusion. In a channel of finite size $L$ perpendicular to the shear plane, the probability to find bacteria at a height $z$ is then found to depend on the applied shear rate through the P\'{e}clet number $\rm Pe = \langle \Zdot_{\text{tot}} \rangle L/\mathcal{D}_{z}$, where $\mathcal{D}_{z}$ is the effective translational diffusion coefficient, see Eqn.~\eqref{eqn:analyticalsolution}. For $\rm Pe \ll 1$, bacteria are uniformly distributed along $z$, while for $\rm Pe \gg 1$, all bacteria accumulate at the boundary $z=L$. The fraction of bacteria accumulating at the boundary as a function of the applied shear rate was determined experimentally and fitted to the theoretical shape given by Eqn.~\eqref{eq:fracBacteriaVsPe} with a single fit parameter $\mathcal{D}_z$, see Fig.~\ref{fig:bacteria}b. Experimental and theoretical results are in good agreement and yield an effective diffusion coefficient of $\mathcal{D}_{z} \approx 220 \mu \rm m^2/s$.

While the work shown here focused on simple flagellar geometries and simple shear flows, the semi-analytical workflow can be readily extended to more complex problems which may not be amenable to fully analytical formulas. In particular, flows encountered by microorganisms in their natural environment are unsteady~\cite{stocker2012}, introducing another characteristic time scale in the problem; the decoupling of time scales between Jeffery effects and rheotactic effects could then prove useful to simplify the analysis of rheotactic drift in these natural environments.

\section{Material and Methods}

\subsection{Bacteria}

Experiments were performed with \textit{Bacillus subtilis} strain GM2938, a kind gift from Dominique Le Coq (Micalis Institute, INRA, Universit\'{e} Paris-Sud). The day before an experiment, cells from an LB-agar plate were resuspended in LB medium and grown overnight in an incubator set at 37\degree C, with a rotation rate of  180 rpm. On the morning of the experiment, 10~$\mu\rm L$ of the overnight bacterial suspension was diluted into 2~mL LB medium (Sigma-Aldrich L3022, Merck, Germany), and bacteria were let grow in the incubator until they reached an optical density at 600~nm of $\rm OD_{600}\approx 0.2$ (Ultrospec 100, Amersham Biosciences, UK). The bacterial suspension was then introduced in the microfluidic chip to form droplets as detailed below.   

\subsection{Microfluidics}

The microfluidic device used was similar to the one described in~\cite{Amselem2016UniversalDroplets}, and consisted in a wide, 2D chamber ($\rm 0.5\times3 \; cm$) connected to two inlets and one outlet. An array of square grooves, or anchors, was etched on the chamber ceiling. The chamber depth was $35\; \mu \rm m$, and the anchors had a side $150 \mu \rm m$ and height $100 \mu \rm m$, see Fig.~\ref{fig:device}a-b for a sketch of the device. All devices were made out of PDMS and sealed on a glass microscopy slide by plasma bonding.

Before the beginning of experiments, microfluidic devices were rendered hydrophobic by surface treatment with Novec 1720 (3M), and prefilled with fluorinated oil (FC-40, 3M) containing 0.5\% (w/w) surfactant (RAN Technologies). Droplets of bacterial suspension were created in two steps: first, an aqueous suspension of bacteria was flowed into the device, and then a gentle flow of oil was applied. This led to the break-up of droplets on the anchors, see~\cite{Amselem2015BreakingCell,Amselem2016UniversalDroplets} for details. A typical snapshot of a droplet formed by break-up is shown on Fig.~\ref{fig:device}c, where individual bacteria can be seen. As a result of the break-up process, droplets were immobilized in their respective anchors and surrounded by oil.

\subsection{Data and Code availability}

All data needed to evaluate the conclusions in the paper are available upon reasonable request until publication. The codes necessary to reproduce the results presented are available on \href{https://github.com/gronteix/rheotaxis_code}{GitHub}.

\section{Extended Data Figures}

\bibliographystyle{unsrt}  
\bibliography{biblio,references}

\begin{thebibliography}{10}

\bibitem{Purcell1977LifeNumber}
E~M Purcell.
\newblock {Life at Low Reynolds Number}.
\newblock 45(June 1976):3--11, 1977.

\bibitem{Lauga2009TheMicroorganisms}
Eric Lauga and Thomas~R. Powers.
\newblock {The hydrodynamics of swimming microorganisms}.
\newblock {\em Reports on Progress in Physics}, 72(9), 2009.

\bibitem{Kantsler2014RheotaxisCells}
Vasily Kantsler, Jörn Dunkel, Martyn Blayney, and Raymond~E Goldstein.
\newblock {Rheotaxis facilitates upstream navigation of mammalian sperm cells}.
\newblock {\em eLife}, 3:1--12, 2014.

\bibitem{kessler1985}
John~O. Kessler.
\newblock {Hydrodynamic focusing of motile algal cells}.
\newblock {\em Nature}, 313(5999):218--220, jan 1985.

\bibitem{garcia2013}
Xabel Garcia, Salima Rafa{\"{i}}, and Philippe Peyla.
\newblock {Light Control of the Flow of Phototactic Microswimmer Suspensions}.
\newblock {\em Physical Review Letters}, 110(13):138106, mar 2013.

\bibitem{marcos2009}
Marcos, Henry~C Fu, Thomas~R Powers, and Roman Stocker.
\newblock {Separation of microscale chiral objects by shear flow}.
\newblock {\em Physical Review Letters}, 102(15):1--4, 2009.

\bibitem{marcos2012}
Marcos, H.~C. Fu, T.~R. Powers, and R.~Stocker.
\newblock {Bacterial rheotaxis}.
\newblock {\em Proceedings of the National Academy of Sciences},
  109(13):4780--4785, 2012.

\bibitem{mathijssen2019}
Arnold~JTM Mathijssen, Nuris Figueroa-Morales, Gaspard Junot, {\'E}ric
  Cl{\'e}ment, Anke Lindner, and Andreas Z{\"o}ttl.
\newblock Oscillatory surface rheotaxis of swimming e. coli bacteria.
\newblock {\em Nature communications}, 10(1):1--12, 2019.

\bibitem{Jing2020Chirality-inducedFlows}
Guangyin Jing, Andreas Z{\"{o}}ttl, Éric Cl{\'{e}}ment, and Anke Lindner.
\newblock {Chirality-induced bacterial rheotaxis in bulk shear flows}.
\newblock {\em Science Advances}, 6(28), 2020.

\bibitem{Jeffery1922TheFlow}
G.~B. Jeffery and Louis Napoleon~George Filon.
\newblock {The Motion of Ellipsoidal Particles Immersed in shear flow}.
\newblock {\em Proc. R. Soc. A}, 102(715), 1922.

\bibitem{ishimoto2020}
Kenta Ishimoto.
\newblock {Helicoidal particles and swimmers in a flow at low Reynolds number}.
\newblock {\em Journal of Fluid Mechanics}, 2020.

\bibitem{Marcos2012BacterialRheotaxis}
{Marcos}, H.~C. Fu, T.~R. Powers, and R.~Stocker.
\newblock {Bacterial rheotaxis}.
\newblock {\em Proceedings of the National Academy of Sciences},
  109(13):4780--4785, 2012.

\bibitem{childress_1981}
Stephen Childress.
\newblock {\em Mechanics of Swimming and Flying}.
\newblock Cambridge Studies in Mathematical Biology. Cambridge University
  Press, 1981.

\bibitem{HildingFaxenDerIst}
{Hilding Faxen}.
\newblock {Der Widerstand gegen die Bewegung einer starren Kugel in einer
  z{\"{a}}hen Fl{\"{u}}ssigkeit, die zwischen zwei parallelen ebenen
  W{\"{a}}nden eingeschlossen ist}.

\bibitem{Curie1894SurMagnetique}
Pierre Curie.
\newblock {Sur la sym{\'{e}}trie dans les ph{\'{e}}nom{\`{e}}nes physiques,
  sym{\'{e}}trie d'un champ {\'{e}}lectrique et d'un champ magn{\'{e}}tique}.
\newblock {\em J. Phys. Theor. Appl.}, 3(1), 1894.

\bibitem{Mathijssen2019OscillatoryBacteria}
Arnold~J.T.M. Mathijssen, Nuris Figueroa-Morales, Gaspard Junot, Éric
  Cl{\'{e}}ment, Anke Lindner, and Andreas Z{\"{o}}ttl.
\newblock {Oscillatory surface rheotaxis of swimming E. coli bacteria}.
\newblock {\em Nature Communications}, 10(1):7--9, 2019.

\bibitem{jeffery1922}
G.~B. Jeffery.
\newblock {The Motion of Ellipsoidal Particles Immersed in a Viscous Fluid}.
\newblock {\em Proceedings of the Royal Society A: Mathematical, Physical and
  Engineering Sciences}, 102(715):161--179, nov 1922.

\bibitem{Leal1971TheFlow}
L.~G. Leal and E.~J. Hinch.
\newblock {The effect of weak Brownian rotations on particles in shear flow}.
\newblock {\em Journal of Fluid Mechanics}, 46(4):685--703, 1971.

\bibitem{Hinch1979RotationFlow}
E.~J. Hinch and L.~G. Leal.
\newblock {Rotation of small non-axisymmetric particles in a simple shear
  flow}.
\newblock {\em Journal of Fluid Mechanics}, 92(3):591--607, 1979.

\bibitem{Dangla2011TrappingEnergy}
Rémi Dangla, Sungyon Lee, and Charles~N. Baroud.
\newblock {Trapping microfluidic drops in wells of surface energy}.
\newblock {\em Physical Review Letters}, 107(12), 2011.

\bibitem{Amselem2016UniversalDroplets}
Gabriel Amselem, Cyprien Guermonprez, Benoît Drogue, Sébastien Michelin, and
  Charles~N. Baroud.
\newblock {Universal microfluidic platform for bioassays in anchored droplets}.
\newblock {\em Lab on a Chip}, 16(21):4200--4211, 2016.

\bibitem{Lee2012}
Sungyon Lee, Fran{\c{c}}ois Gallaire, and Charles~N. Baroud.
\newblock {Interface-induced recirculation within a stationary microfluidic
  drop}.
\newblock {\em Soft Matter}, 8(41):10750, 2012.

\bibitem{ling2016}
Yue Ling, Jose~Maria Fullana, St{\'{e}}phane Popinet, and Christophe Josserand.
\newblock {Droplet migration in a Hele-Shaw cell: Effect of the lubrication
  film on the droplet dynamics}.
\newblock {\em Physics of Fluids}, 28(6), 2016.

\bibitem{berg1993random}
Howard~C. Berg.
\newblock {\em {Random Walks in Biology}}.
\newblock Princeton University Press, nov 1993.

\bibitem{Saragosti2012ModelingChemotaxis}
Jonathan Saragosti, Pascal Silberzan, and Axel Buguin.
\newblock {Modeling E. coli tumbles by rotational diffusion. implications for
  chemotaxis}.
\newblock {\em PLoS ONE}, 7(4), 2012.

\bibitem{stocker2012}
Roman Stocker.
\newblock {Marine Microbes See a Sea of Gradients}.
\newblock {\em Science}, 338(6107):628--633, 2012.

\bibitem{Amselem2015BreakingCell}
Gabriel Amselem, P.~T. Brun, François Gallaire, and Charles~N. Baroud.
\newblock {Breaking Anchored Droplets in a Microfluidic Hele-Shaw Cell}.
\newblock {\em Physical Review Applied}, 3(5):1--5, 2015.

\end{thebibliography}
%\printbibliography

%%%%%%%%%%%

\end{document}